\documentclass[11pt,a4paper]{article}
\pdfoutput=1
\usepackage{jheppub} 

\usepackage{bm}
\usepackage{slashed}

\preprint{TU-1325}

\title{Environment-Dependent Annihilation of Axion-Coupled Higgsino Dark Matter}

\author[a,b]{Kwang Sik Jeong,}
\author[c,d]{Fuminobu Takahashi\,}

\affiliation[a]{Department of Physics, Pusan National University, Busan 46241, Korea}
\affiliation[b]{School of Physics, Korea Institute for Advanced Study, Seoul 02455, Korea}
\affiliation[c]{Department of Physics, Tohoku University, Sendai, Miyagi 980-8578, Japan}
\affiliation[d]{Kavli IPMU (WPI), UTIAS, University of Tokyo, Kashiwa 277-8583, Japan}

\emailAdd{ksjeong@pusan.ac.kr}
\emailAdd{fumi@tohoku.ac.jp}

\abstract{
We study environment-dependent annihilation of quasi-Dirac Higgsino dark matter coupled to an axion-like particle with a mass slightly below the Higgsino-pair threshold. Axion-mediated annihilation becomes more important relative to electroweak annihilation in the Sun and Galactic halo than during cosmological reannihilation following late nonthermal production. This separation arises from both the low-velocity enhancement of the subthreshold propagator and the change from a coannihilating Higgsino multiplet to the lightest neutral state.  In solar capture–annihilation equilibrium, a dominant annihilation channel into massless dark photons suppresses the high-energy neutrino signal while leaving capture efficient. Higgsinos can consequently account for a few percent of the total dark matter abundance while satisfying solar-neutrino constraints and retaining an appreciable terrestrial inelastic-scattering rate, with implications for the LZ high-recoil candidate. In the Galactic halo, the enhanced cross section instead produces correlated continuum and line gamma-ray signals, with the lines potentially close to the present sensitivity for a cuspy Galactic dark-matter profile. We discuss supersymmetric realizations based on generalized mirage mediation and hidden strong dynamics, as well as an extended Higgs-sector alternative. The dark photons also contribute to dark radiation, providing a complementary cosmological probe.}


\begin{document}
\maketitle
\flushbottom

\section{Introduction}
\label{sec:introduction}

A quasi-Dirac Higgsino with a small neutral-state mass splitting is
an interesting target for inelastic dark-matter searches~\cite{Tucker-Smith:2001myb,Nagata:2014wma,Bramante:2016rdh,Krall:2017xij,Graham:2024syw}. The
off-diagonal $Z$ interaction allows the lightest neutral Higgsino to
up-scatter into the heavier neutral state on nuclei.  The same
interaction, however, also leads to efficient capture in the Sun~\cite{Press:1985ug,Silk:1985ax,Griest:1986yu,Gould:1987ir,Nussinov:2009ft,Menon:2009qj}.
In the range of neutral-state mass splittings for which inelastic
solar capture remains efficient, high-energy solar-neutrino searches~\cite{IceCube:2016yoy,IceCube:2016dgk,IceCube:2025fcu} 
place stringent constraints on quasi-Dirac Higgsino dark matter~\cite{Catena:2018vzc,Pospelov:2026ewn}.
Terrestrial inelastic scattering and solar-neutrino searches
therefore probe closely related regions of the quasi-Dirac Higgsino
parameter space.
This connection has become particularly relevant after the LZ collaboration reported a single nuclear-recoil candidate at $248\,$keV~\cite{LZ:2026axp}: its interpretation in terms of inelastic Higgsino-like dark matter~\cite{Fan:2026hzw,Freese:2026sga,Wu:2026nhi,Yin:2026jnn,Visinelli:2026kgt,DiMauro:2026ldr,Smirnov:2026aqk,Du:2026guj,Langhoff:2026ujr,Bisal:2026LZ} is strongly constrained by solar capture~\cite{Pospelov:2026ewn,Bose:2026ndd,DiMauro:2026dqp,Nguyen:2026lui} as well as by the high-energy sideband~\cite{Rodd:2026tyn}. 
See Refs.~\cite{Aghaie:2026vsu,Lee:2026zbr} for interpretations involving invisible neutron
decay and annihilation. 
Boosted dark-matter interpretations of the high-recoil event have
also been proposed~\cite{Liang:2026LZ,Alhazmi:2026LZ,
Heikinheimo:2026LZ,Jho:2026LZ}.

A simple way to weaken the solar-neutrino constraint is to dilute the
Higgsino abundance.  We define the Higgsino fraction by
\begin{equation}
    \xi_\chi \equiv \frac{\Omega_\chi}{\Omega_{\rm DM}},
\end{equation}
where the observed dark-matter abundance is
$\Omega_{\rm DM}h^2 \simeq 0.12$ and $h$ is the reduced Hubble constant.  As long as capture and annihilation
remain in equilibrium, the solar-neutrino flux scales approximately
as $\xi_\chi$.  Such dilution, however, simultaneously suppresses
other signals of the same quasi-Dirac Higgsino. At fixed Higgsino
mass and neutral-state splitting, the terrestrial
inelastic-scattering rate scales linearly with $\xi_\chi$, while the
standard Galactic-halo annihilation signal scales as $\xi_\chi^2$.
Interestingly, this correlated suppression can be avoided if the
annihilation dynamics depend on the environment.  The solar-neutrino
signal can then be suppressed after capture without requiring a
comparably strong dilution of the Higgsino abundance.
Other ways to reconcile inelastic electroweak dark matter with the solar-neutrino bound include an asymmetric dark-matter population~\cite{Nagata:2026pbj}, singlet--doublet mixing that suppresses both capture and annihilation~\cite{Lee:2026jxl}, suppressed annihilation of the captured population~\cite{DiMauro:2026dqp,Qi:2026vyp}, isospin-violating couplings that suppress capture~\cite{Okada:2026upm}, a very heavy Higgsino~\cite{Langhoff:2026ujr}, and an environment-dependent mass splitting that is enlarged inside the Sun~\cite{Girmohanta:2026qzo}.
In contrast to the last proposal, which suppresses solar capture itself, the mechanism considered here keeps capture efficient and makes the annihilation dynamics environment dependent.

We realize this environment-dependent annihilation by coupling the
Higgsino to an axion-like pseudo-Nambu--Goldstone boson (pNGB) whose
mass lies slightly below the Higgsino-pair threshold.  The resulting
environmental dependence has two distinct origins.  First, the
subthreshold mediator makes axion-mediated annihilation increasingly
important at low velocities.  Second, the composition of the
annihilating Higgsino population differs between cosmological
reannihilation and low-temperature environments.  During
cosmological reannihilation following late nonthermal production,
the nearly degenerate neutral and charged Higgsino states remain
populated and participate in coannihilation.  By contrast, after
thermalization in the Sun the annihilating population is dominated
by the lightest neutral Higgsino, as it is in the present Galactic
halo.  
The combination of these velocity and population effects can
therefore make axion-mediated annihilation substantially more
important in low-temperature environments than during cosmological
reannihilation.  Importantly, the population effect alone can
provide an order-of-magnitude environmental separation,
independently of the near-threshold velocity enhancement.

The resulting environmental enhancement has qualitatively different
consequences in the Sun and in the Galactic halo.
 In the Sun, once
capture--annihilation equilibrium is established, the total
annihilation rate is fixed by the capture rate and is therefore
insensitive to a further enhancement of the annihilation cross
section.  The enhanced axion-mediated channel instead changes the
composition of the annihilation final states.  If it predominantly
feeds neutrino-poor hidden states, the fraction of annihilations
producing high-energy neutrinos is reduced, thereby suppressing the
observable solar-neutrino signal.  In the Galactic halo, by contrast,
there is no capture--annihilation equilibrium, and the annihilation
rate depends directly on the low-velocity cross section.  The same
environmental enhancement can therefore amplify the halo signal, so
that even subdominant visible decay modes of the mediator can give
rise to continuum and monochromatic gamma-ray signals.

We consider a generic nonthermal cosmological history in which a
sufficiently large Higgsino population is injected at late times and
subsequently undergoes reannihilation, without assuming a specific
origin for the nonthermal population.
We then examine
whether a sizable Higgsino fraction can remain compatible with
solar-neutrino constraints while yielding potentially observable
gamma-ray signatures from the Galactic halo.

A sizable Higgsino fraction also has direct implications for
terrestrial inelastic searches.  For a given Higgsino mass and
neutral-state splitting, the $Z$-mediated inelastic-scattering cross
section is fixed by the electroweak quantum numbers of the Higgsino,
while the kinematics of the endothermic transition makes the event
rate sensitive to the high-velocity tail of the Galactic Higgsino
velocity distribution.  For a fixed velocity distribution, and
assuming that the local Higgsino fraction follows its cosmological
abundance fraction, the terrestrial event rate scales linearly with
$\xi_\chi$.  Owing to the sizable electroweak scattering cross
section, terrestrial inelastic searches can remain sensitive even
when the Higgsino constitutes only a subdominant component of dark
matter, provided that the high-velocity tail contains a sufficient
population above the inelastic threshold.  By relaxing the
solar-neutrino bound on $\xi_\chi$ without relying on strong abundance
dilution, the environment-dependent annihilation mechanism can
therefore preserve a substantially larger terrestrial
inelastic-scattering rate than in the standard Higgsino dark matter scenario.

Finally, we discuss a supersymmetric realization based on generalized
mirage mediation in KKLT-type compactification.  This framework
naturally accommodates a heavy gravitino, whose late decay can
nonthermally produce Higgsinos that subsequently undergo
reannihilation.  At the same time, the interplay between modulus and
anomaly mediation allows a cancellation between the bino- and
wino-induced contributions to the neutral-Higgsino mass splitting.
A quasi-Dirac Higgsino can therefore be realized without decoupling
the electroweak gauginos at an extremely high scale.  Generalized
mirage mediation thus provides a concrete framework in which both the
nonthermal cosmological history and the quasi-Dirac Higgsino spectrum
considered here can arise, while the main phenomenological results do
not rely on this particular realization.

This paper is organized as follows.  
In Sec.~\ref{sec:environment}, we specify the quasi-Dirac Higgsino spectrum and the axion mediator and derive the environmental enhancement relating the cosmological, solar, and Galactic-halo annihilation rates.
In Sec.~\ref{sec:axion-couplings}, we discuss the axion couplings to Standard Model fermions and to visible and dark gauge bosons, which determine the decay pattern of the mediator.
In Sec.~\ref{sec:combined-phenomenology}, we connect the environmental enhancement to the nonthermal Higgsino abundance and set up the framework for solar-neutrino and Galactic gamma-ray signals.
The resulting constraints and signatures, including terrestrial inelastic scattering, are presented in Sec.~\ref{sec:phenomenological-implications}.
A supersymmetric realization based on generalized mirage mediation is discussed in Sec.~\ref{sec:susy-realization}, and Sec.~\ref{sec:conclusions} is devoted to conclusions.

\section{Environment-dependent annihilation}
\label{sec:environment}

The neutral-state mass splitting of a quasi-Dirac Higgsino plays a
central role in its solar phenomenology.  The off-diagonal $Z$
interaction permits inelastic up-scattering of the lightest neutral
Higgsino into the heavier neutral state on solar nuclei, while the
mass splitting controls the kinematic accessibility of this process.
As the splitting increases, inelastic solar capture eventually becomes
suppressed, thereby weakening the associated solar-neutrino
constraint.  We focus instead on the regime in which solar capture
remains efficient.  In this regime, suppressing the solar-neutrino
signal requires modifying the annihilation dynamics after capture,
rather than relying on kinematic suppression of the capture rate.

We introduce an additional annihilation channel mediated by an
axion-like pNGB whose mass lies slightly below the Higgsino-pair
threshold.  Its importance relative to the standard electroweak
channels depends on the environment through two distinct effects.
First, the subthreshold mediator enhances axion-mediated annihilation
toward low velocities.  Second, the composition of the annihilating
Higgsino population differs between cosmological reannihilation
following late nonthermal production and low-temperature environments
such as the Sun and the present Galactic halo.  During reannihilation,
the nearly degenerate neutral and charged Higgsino states participate
in coannihilation, whereas in the low-temperature environments the
annihilating population is dominated by the lightest neutral state.

Together, these effects can make axion-mediated annihilation
substantially more important in the Sun and the Galactic halo than
during cosmological reannihilation.  We first specify the quasi-Dirac
Higgsino spectrum and the axion-mediated interaction, and then analyze
the velocity and population effects separately before combining them
into a quantitative relation among the cosmological, solar, and
Galactic-halo annihilation rates.

\subsection{Quasi-Dirac Higgsinos and the axion mediator}
\label{subsec:quasi-dirac-solar}

We consider Higgsino-like neutralino dark matter in the quasi-Dirac
regime.  We denote the lightest neutralino by
$\chi\equiv\chi_1^0$, the heavier Higgsino-like neutralino by
$\chi_2^0$, and the Higgsino-like chargino by $\chi^\pm$.  The
spectrum of interest is characterized by the hierarchy
\begin{equation}
    \Delta m_0
    \equiv m_{\chi_2^0}-m_\chi
    \ll
    \Delta m_\pm
    \equiv m_{\chi^\pm}-m_\chi
    \ll m_\chi .
    \label{eq:higgsino-hierarchy}
\end{equation}
For a Higgsino-like spectrum, electroweak radiative corrections
naturally generate a charged--neutral splitting at the sub-GeV
scale, whereas the neutral splitting is more model-dependent and
arises from interactions that split the otherwise Dirac neutral
Higgsino pair~\cite{Chun:2016cnm}.  We are particularly interested in the quasi-Dirac
regime in which $\Delta m_0$ is much smaller than $\Delta m_\pm$ and
lies in the range where inelastic scattering remains kinematically
relevant for both terrestrial searches and solar capture. 
We treat $\Delta m_0$ as a low-energy parameter, postponing its
ultraviolet realization to a later discussion.  The results derived
below are insensitive to the particular mechanism generating it.

We introduce an additional annihilation channel mediated by an
axion-like pseudo-Nambu--Goldstone boson, denoted by $\phi$.  An
axion mediator is particularly well suited for our purpose.  Its mass
is protected by the approximate shift symmetry and can therefore
remain light without being destabilized by radiative corrections.
Moreover, its couplings to fermions are controlled by the associated
symmetry charges, while its couplings to gauge bosons are determined
by mixed anomalies.  In particular, the axion can couple to
hidden-sector particles, providing annihilation channels with a
suppressed high-energy neutrino yield.

A simple realization of such a hidden channel is a pair of massless
dark photons $A'$ associated with an unbroken dark gauge symmetry.
We assume negligible kinetic mixing between $U(1)_D$ and hypercharge.
The coupling of $\phi$ to dark photons can arise naturally from a
mixed anomaly, leading to the annihilation channel
\begin{equation}
    \chi\chi \;\to\; \phi^* \;\to\; A'A' .
    \label{eq:annihilation-dark-photon}
\end{equation}
We use this channel as our primary hidden-sector example below,
although the environment-dependent annihilation mechanism does not
rely on dark photons as the final state.

\subsection{Subthreshold axion-mediated annihilation}
\label{subsec:threshold-annihilation}

We first quantify the velocity dependence associated with the
subthreshold mediator.  We parameterize the distance of the pole
below the $\chi\chi$ threshold by
\begin{equation}
    \Delta_\phi
    \equiv
    2m_\chi-m_\phi >0 .
    \label{eq:threshold-detuning}
\end{equation}
For $\Delta_\phi\ll m_\chi$, a nonrelativistic Higgsino pair has
$\sqrt{s}\simeq2m_\chi+m_\chi v_{\rm rel}^2/4$.  The pseudoscalar
coupling of the axion to the Higgsino leads to $s$-wave annihilation,
so that the dominant velocity dependence near threshold arises from
the mediator propagator.  Up to factors that vary slowly near
threshold, the annihilation rate therefore behaves as
\begin{equation}
    \sigma_\phi v_{\rm rel}
    \propto
    \left[
        \left(
            \Delta_\phi
            +\frac{m_\chi v_{\rm rel}^2}{4}
        \right)^2
        +\frac{\Gamma_\phi^2}{4}
    \right]^{-1},
    \label{eq:threshold-velocity-dependence}
\end{equation}
where $\Gamma_\phi$ is the total width of the mediator.  Since the
pole lies below threshold, increasing the relative kinetic energy
moves the center-of-mass energy farther away from the pole.  The
annihilation rate therefore increases toward lower velocities~\cite{Ibe:2008ye}.

The velocity dependence in
Eq.~\eqref{eq:threshold-velocity-dependence} distinguishes
cosmological reannihilation from low-velocity environments such as
the Sun and the present Galactic halo.  During cosmological
reannihilation, the Higgsinos sample a finite thermal velocity
distribution.  The associated kinetic energy shifts the
center-of-mass energy farther away from the subthreshold pole and
reduces the thermally averaged enhancement.  By contrast, for
Higgsinos thermalized in the solar core and for Higgsinos in the
present Galactic halo, the kinetic contribution
$m_\chi v_{\rm rel}^2/4$ is negligible compared with
$\Delta_\phi$.  Both environments therefore probe the effectively
zero-velocity limit.

The finite mediator width regulates the enhancement as the pole
approaches threshold.  The simple analytic relation derived below is
most transparent in the regime
\begin{equation}
    \Gamma_\phi\lesssim\Delta_\phi\ll m_\chi,
    \label{eq:width-regime}
\end{equation}
and further simplifies in the narrow-width limit
$\Gamma_\phi\ll\Delta_\phi$.  Importantly, a sizable environmental
enhancement does not require an extremely small fractional detuning
$\Delta_\phi/m_\chi$.  Its dependence on $\Delta_\phi$ is quantified
in Sec.~\ref{subsec:environmental-enhancement}.

\subsection{Higgsino populations in different environments}
\label{subsec:higgsino-populations}

The velocity dependence discussed in the previous subsection is only
one source of environmental separation.  A second effect arises from
the different compositions of the annihilating Higgsino population
during cosmological reannihilation and in low-temperature
environments such as the Sun and the present Galactic halo.

We focus on a cosmological history in which late nonthermal
production of Higgsinos is followed by reannihilation, as can arise
from the late decay of moduli, gravitinos, saxions, or other
long-lived states.  We denote by $T_{\rm ann}$ the plasma temperature
at which reannihilation effectively completes and consider the
nonrelativistic regime
\begin{equation}
    \Delta m_0
    \ll
    \Delta m_\pm
    \lesssim
    T_{\rm ann}
    \ll
    m_\chi .
    \label{eq:annihilation-temperature-hierarchy}
\end{equation}
For a Higgsino-like spectrum, electroweak radiative corrections
naturally generate a charged--neutral splitting at the sub-GeV
scale.  In this regime, the charged Higgsinos remain thermally
populated, although their abundances can be Boltzmann suppressed.
Electroweak scattering and conversion processes maintain kinetic
equilibrium with the thermal plasma and chemical equilibrium within
the Higgsino multiplet.  Neglecting the much smaller neutral
splitting in the equilibrium abundances, the corresponding fractions
are approximately 
\begin{equation}
    r_\chi \simeq r_2 \simeq
    \frac{1}{2(1+b_\pm)},
    \qquad
    r_\pm \simeq
    \frac{b_\pm}{2(1+b_\pm)},
    \qquad
    b_\pm \equiv
    \exp\left(-\frac{\Delta m_\pm}{T_{\rm ann}}\right),
    \label{eq:higgsino-fractions}
\end{equation}
where $b_\pm$ is the charged-Higgsino Boltzmann factor and $r_\pm$
denotes the equilibrium fraction of each charged Higgsino state.

To determine how the individual Higgsino states contribute to
axion-mediated annihilation, we note that, owing to its
Nambu--Goldstone nature, the axion inherits its coupling to the
Higgsinos from the Higgsino mass term.  In the pure-Higgsino limit,
this coupling is diagonal in the neutral Higgsino mass eigenstates,
so that the off-diagonal $\phi\chi\chi_2^0$ coupling vanishes at
leading order, up to corrections from electroweak-gaugino mixing.
The effective axion-mediated annihilation rate during cosmological
reannihilation is therefore~\cite{Griest:1990kh}
\begin{equation}
    \langle\sigma v\rangle^{\rm cos}_{\phi,{\rm eff}}
    =
    r_\chi^2
    \langle\sigma v\rangle^{\rm cos}_{\phi,\chi\chi}
    +
    r_2^2
    \langle\sigma v\rangle^{\rm cos}_{\phi,22}
    +
    2r_\pm^2
    \langle\sigma v\rangle^{\rm cos}_{\phi,+-}.
    \label{eq:cos-axion-effective}
\end{equation}
The charged-Higgsino pair lies farther from the mediator pole because
its threshold is higher than the $\chi\chi$ threshold.  Its effective
detuning is approximately
\begin{equation}
    \Delta_{\phi,\pm}
    \simeq
    \Delta_\phi+2\Delta m_\pm ,
    \label{eq:charged-detuning}
\end{equation}
so that the charged channel receives a smaller near-threshold
enhancement than the neutral channels.

By the time Galactic-halo Higgsinos reach the Sun, the heavier states populated during cosmological reannihilation are assumed to have decayed, leaving an incident population dominated by the lightest neutral state $\chi$. Inelastic scattering during solar capture can nevertheless excite $\chi$ to $\chi_2^0$. After capture and thermalization, the characteristic temperature in the solar core satisfies
\begin{equation}
T_\odot={\cal O}({\rm keV})
\ll
\Delta m_0,\Delta m_\pm ,
\label{eq}
\end{equation}
so that thermal repopulation of the heavier states is strongly suppressed. As the captured Higgsinos relax to the lightest neutral state, the annihilating population becomes dominated by $\chi$. The effective axion-mediated annihilation rate in the Sun is therefore well approximated by
%
\begin{equation}
    \langle\sigma v\rangle^\odot_{\phi,{\rm eff}}
    \simeq
    \langle\sigma v\rangle^\odot_{\phi,\chi\chi}.
    \label{eq:solar-axion-effective}
\end{equation}
Thus, unlike during cosmological reannihilation, the heavier neutral
and charged Higgsino states do not contribute appreciably to the
annihilating population in the Sun.

The present Galactic halo has the same relevant annihilating
population, with the heavier Higgsino states negligibly populated.
Present-day halo annihilation therefore also involves essentially
only the lightest neutral state,
\begin{equation}
    \langle\sigma v\rangle^{\rm halo}_{\phi,{\rm eff}}
    \simeq
    \langle\sigma v\rangle^{\rm halo}_{\phi,\chi\chi}.
    \label{eq:halo-axion-effective}
\end{equation}
Together with the effectively zero-velocity behavior established in
Sec.~\ref{subsec:threshold-annihilation}, this shows that the Sun and
the Galactic halo probe essentially the same axion-mediated
annihilation cross section.

The population effect therefore acts in the same direction as the
velocity effect discussed in the previous subsection.  During
cosmological reannihilation, the full nearly degenerate Higgsino
multiplet participates, including the charged channel whose
threshold lies farther from the subthreshold mediator pole.  In the
Sun and the present Galactic halo, by contrast, the annihilating
population is dominated by the lightest neutral state.  The relative
importance of axion-mediated annihilation can consequently be
substantially larger in these low-temperature environments than
during cosmological reannihilation.  We combine the population and
velocity effects quantitatively in the next subsection.

\subsection{Environmental enhancement}
\label{subsec:environmental-enhancement}

We now combine the velocity and population effects discussed above
and quantify the annihilation enhancement in the different
environments.  During cosmological reannihilation, we define the
total annihilation rate relative to the electroweak contribution by
\begin{equation}
    R_{\rm cos}
    \equiv
    \frac{
        \langle\sigma v\rangle^{\rm cos}_{{\rm tot},{\rm eff}}
    }{
        \langle\sigma v\rangle^{\rm cos}_{{\rm EW},{\rm eff}}
    }.
    \label{eq:Rcos-general}
\end{equation}
For Higgsinos thermalized in the Sun, the corresponding quantity is
\begin{equation}
    R_\odot
    \equiv 
    \frac{
        \langle\sigma v\rangle^\odot_{{\rm tot},\chi\chi}
    }{
        \langle\sigma v\rangle^\odot_{{\rm EW},\chi\chi}
    }.
    \label{eq:Rsun-general}
\end{equation}
Here the total rates include interference between
amplitudes contributing to the same final state.

It is useful to define the environmental enhancement factor
\begin{equation}
    K_{\rm env}
    \equiv
    \frac{
        \langle\sigma v\rangle^\odot_{\phi,\chi\chi}
    }{
        \langle\sigma v\rangle^{\rm cos}_{\phi,{\rm eff}}
    }
    \frac{
        \langle\sigma v\rangle^{\rm cos}_{{\rm EW},{\rm eff}}
    }{
        \langle\sigma v\rangle^\odot_{{\rm EW},\chi\chi}
    }.
    \label{eq:Kenv-definition}
\end{equation}
Thus $K_{\rm env}$ directly measures the enhancement of the
axion-to-electroweak annihilation ratio between cosmological
reannihilation and the low-velocity solar environment.  
In the parameter region of interest, the axion-mediated annihilation 
rate is dominated by the dark-photon channel, which has no corresponding
standard Higgsino amplitude.  Interference is therefore restricted
to subdominant visible channels and gives only a small correction to
the total annihilation rate, even when it is sizable
relative to the axion-mediated contribution in an individual visible
channel.  We consequently have 
\begin{equation}
    R_\odot
    \simeq
    1+K_{\rm env}\left(R_{\rm cos}-1\right).
    \label{eq:Rsun-Kenv}
\end{equation}
Interference can nevertheless be important for
individual visible final states and will be included when evaluating
the corresponding observables.
As shown below, the enhancement encoded in $K_{\rm env}$ has two
distinct origins.
One is the lower characteristic velocity of Higgsinos in the Sun,
and the other is the change in the annihilating population from the
full Higgsino multiplet during cosmological reannihilation to the
lightest neutral state in the Sun.

The present Galactic halo provides another low-velocity environment.
As in the Sun, its annihilating Higgsino population consists almost
entirely of $\chi$, and both environments probe approximately the same
zero-velocity annihilation cross sections. We therefore define
%
\begin{equation}
    R_{\rm halo}
    \equiv 
    \frac{
        \langle\sigma v\rangle^{\rm halo}_{{\rm tot},\chi\chi}
    }{
        \langle\sigma v\rangle^{\rm halo}_{{\rm EW},\chi\chi}
    }.
    \label{eq:Rhalo-general}
\end{equation}
The results of
Secs.~\ref{subsec:threshold-annihilation} and
\ref{subsec:higgsino-populations} then imply
\begin{equation}
    R_{\rm halo}
    \simeq
    R_\odot ,
    \label{eq:Rhalo-Rsun}
\end{equation}
up to small corrections from the residual velocity dependence of
the annihilation cross sections.

We first quantify the change in the electroweak annihilation rate
associated with the different Higgsino populations.  In the
degenerate pure-Higgsino limit, the leading-order effective
electroweak coannihilation rate during cosmological reannihilation is~\cite{Jungman:1995df,Nihei:2002ij,Beneke:2012tg}
\begin{equation}
    \langle\sigma v\rangle^{\rm cos}_{{\rm EW},{\rm eff}}
    \simeq
    \frac{g_2^4}{512\pi m_\chi^2}
    \left(
        21+3\tan^2\theta_W+11\tan^4\theta_W
    \right),
    \label{eq:EW-cosmological}
\end{equation}
whereas in the low-temperature Sun, where only the lightest neutral
Higgsino is populated, the corresponding leading-order low-velocity
annihilation rate is
\begin{equation}
    \langle\sigma v\rangle^\odot_{{\rm EW},\chi\chi}
    \simeq
    \frac{g_2^4}{128\pi m_\chi^2}
    \left(
        1+\frac{1}{2\cos^4\theta_W}
    \right).
    \label{eq:EW-solar}
\end{equation}
Their ratio is
\begin{equation}
    \frac{
        \langle\sigma v\rangle^{\rm cos}_{{\rm EW},{\rm eff}}
    }{
        \langle\sigma v\rangle^\odot_{{\rm EW},\chi\chi}
    }
    \simeq 3.1,
    \label{eq:EW-environment-ratio}
\end{equation}
where we used $\sin^2\theta_W\simeq0.231$.
The numerical value corresponds to the degenerate-Higgsino limit,
$b_\pm\to1$.  At leading order, this ratio is independent of
$m_\chi$ in the pure-Higgsino limit.  Away from the degenerate
limit, the effective electroweak coannihilation rate acquires an
additional dependence on $b_\pm$.  Thus, in the degenerate limit,
the difference
in the Higgsino population alone makes the effective
electroweak annihilation rate during cosmological reannihilation
about three times larger than the low-velocity $\chi\chi$
annihilation rate.

We next quantify the velocity dependence of the axion-mediated
contribution.  For a Higgsino pair whose threshold lies a distance
$d$ above the mediator pole, the thermally averaged annihilation
rate has the parametric dependence
\begin{equation}
    \langle\sigma v\rangle_d
    \propto
    \int_0^\infty dx\,
    \frac{x^{1/2}e^{-x}}
    {(d+xT_{\rm ann})^2+\Gamma_\phi^2/4},
    \label{eq:thermal-average-threshold}
\end{equation}
where $x\equiv m_\chi v_{\rm rel}^2/(4T_{\rm ann})$ is the
dimensionless relative kinetic energy.  Neglecting the much smaller
neutral-Higgsino splitting, the detuning is
$d\simeq\Delta_\phi$ for the neutral channels and
$d\simeq\Delta_\phi+2\Delta m_\pm$ for the charged channel.  During
cosmological reannihilation, the thermal distribution samples finite
relative kinetic energies that move the annihilating pair farther
from the subthreshold pole, with the charged channel shifted still
farther because of its higher threshold.

For the Sun and the present Galactic halo, by contrast, the kinetic
contribution is negligible compared with $\Delta_\phi$, as discussed
in Sec.~\ref{subsec:threshold-annihilation}.  The axion-mediated
annihilation rate of the lightest neutral Higgsino is therefore well
approximated by
\begin{equation}
    \langle\sigma v\rangle^\odot_{\phi,\chi\chi}
    \simeq
    \langle\sigma v\rangle^{\rm halo}_{\phi,\chi\chi}
    \propto
    \frac{1}
    {\Delta_\phi^2+\Gamma_\phi^2/4}.
    \label{eq:low-velocity-axion-rate}
\end{equation}
The axion-mediated annihilation rate is therefore enhanced in these
low-velocity environments relative to its thermally averaged value
during cosmological reannihilation.

The two effects entering $K_{\rm env}$ can be separated, to a good
approximation, as
\begin{equation}
    K_{\rm env}
    \simeq
    K_0 F_v ,
    \label{eq:Kenv-factorization}
\end{equation}
where $K_0$ captures the population-induced change in the relative
importance of axion-mediated annihilation in the absence of the
velocity-dependent subthreshold enhancement, while $F_v$ accounts
for the additional enhancement associated with the lower Higgsino
velocities in the Sun.

The population-induced factor $K_0$ follows directly from the
Higgsino fractions in Eq.~\eqref{eq:higgsino-fractions}.  In the
limit $\Delta_\phi\gg T_{\rm ann},\Delta m_\pm$, the propagator
becomes insensitive to both the thermal kinetic energy and the small
differences among the Higgsino thresholds.  The remaining
environmental separation is then determined entirely by the
different Higgsino populations entering the axion-mediated and
electroweak annihilation rates.  Using $r_2\simeq r_\chi$, we obtain
\begin{equation}
    K_0
    \simeq
    \frac{1}{2r_\chi^2+2r_\pm^2}
    \frac{
        \langle\sigma v\rangle^{\rm cos}_{{\rm EW},{\rm eff}}
    }{
        \langle\sigma v\rangle^\odot_{{\rm EW},\chi\chi}
    }.
    \label{eq:K0-population}
\end{equation}
In the degenerate limit $b_\pm\to1$, using
Eq.~\eqref{eq:EW-environment-ratio} gives
\begin{equation}
    K_0 \to 12.4 .
\end{equation}
Thus, in this limit, the change in the annihilating Higgsino
population alone enhances the relative importance of axion-mediated
annihilation by more than an order of magnitude in the
low-temperature environment.
This population-induced enhancement does not rely on proximity to
the mediator pole and persists even when the velocity-dependent
subthreshold enhancement becomes negligible.  It therefore provides
a robust source of environmental separation without requiring
$\Delta_\phi/T_{\rm ann}\ll1$.
Away from the degenerate limit, both the Higgsino population
fractions and the effective electroweak coannihilation rate depend
on $b_\pm$ and should be treated consistently.

The additional velocity enhancement can be understood analytically
in the narrow-width regime $\Gamma_\phi\ll\Delta_\phi$.  For
$T_{\rm ann}/\Delta_\phi\ll1$, expanding the propagator in
Eq.~\eqref{eq:thermal-average-threshold} and using
$\langle x\rangle=3/2$ gives a thermal suppression
$1-3T_{\rm ann}/d+\cdots$ for a channel with detuning $d$.
The corresponding low-velocity enhancement for the neutral channel
is therefore
\begin{equation}
    F_v
    \simeq
    1+
    3\frac{T_{\rm ann}}{\Delta_\phi}
    +
    {\cal O}\left(
        \frac{T_{\rm ann}^2}{\Delta_\phi^2}
    \right).
    \label{eq:Fv-expansion}
\end{equation}
The full thermal average, including the charged-Higgsino contribution
and its shifted threshold, can be incorporated into the combined
environmental enhancement by an effective coefficient of
approximately $3.3$ over the parameter range considered below.

The leading dependence on $m_\chi$ drops out of the environmental
enhancement.  For a nonrelativistic thermal population,
$v_{\rm rel}^2\sim T_{\rm ann}/m_\chi$, so that the kinetic-energy
shift $m_\chi v_{\rm rel}^2/4$ entering the near-threshold
propagator is controlled by $T_{\rm ann}$ rather than by $m_\chi$.
The electroweak-rate ratio entering $K_0$ is likewise independent of
$m_\chi$ at leading order in the pure-Higgsino limit.

Combining the population and velocity effects gives a simple
approximate relation among the cosmological, solar, and Galactic-halo
annihilation rates.  In the nearly degenerate Higgsino and
narrow-width limits, and neglecting the subleading
interference corrections to the total annihilation rates, a useful
approximation is
\begin{equation}
    R_{\rm halo}
    \simeq
    R_\odot
    \simeq
    1+
    12.4
    \left(
        1+
        3.3\,\frac{T_{\rm ann}}{\Delta_\phi}
    \right)
    \left(R_{\rm cos}-1\right).
    \label{eq:environment-final}
\end{equation}
This relation applies in the near-threshold regime
$\Delta_\phi\ll m_\chi$ and the narrow-width regime
$\Gamma_\phi\ll\Delta_\phi$, together with the hierarchy in
Eq.~\eqref{eq:annihilation-temperature-hierarchy}.
Away from the degenerate limit, the factor $12.4$ is replaced by
$K_0$, with both the Higgsino population fractions and the effective
electroweak coannihilation rate evaluated at the corresponding
$b_\pm$.

The final relation summarizes the two sources of environmental
enhancement.  The change in the annihilating Higgsino population
alone gives an order-of-magnitude enhancement,
$K_0\simeq12.4$ in the degenerate limit, while the subthreshold
mediator provides an additional velocity-dependent enhancement
controlled primarily by $\Delta_\phi/T_{\rm ann}$.  Importantly,
the former does not rely on near-threshold velocity enhancement.
A sizable environmental separation can therefore persist even for
$\Delta_\phi\gtrsim T_{\rm ann}$, without requiring
$\Delta_\phi/T_{\rm ann}\ll1$.  Since the leading explicit
dependence on $m_\chi$ drops out of $K_{\rm env}$, the mechanism can
operate over a broad range of Higgsino masses without requiring an
extremely small fractional detuning $\Delta_\phi/m_\chi$.


\section{Axion couplings and decay pattern}
\label{sec:axion-couplings}

The environmental enhancement derived in
Sec.~\ref{sec:environment} can suppress the solar-neutrino signal
only if the enhanced axion-mediated channel has a sufficiently small
high-energy neutrino yield.  The resulting constraint therefore
depends not only on the relative importance of axion-mediated
annihilation in the Sun, but also on the decay pattern of the
mediator.  An axion-like pNGB is particularly well suited for this
purpose: it can transfer most of the annihilation energy into a
hidden gauge sector through anomaly-induced couplings, while its
unavoidable visible couplings are closely tied to the underlying
symmetry structure.  We denote the spontaneously broken global
symmetry associated with the pNGB by $U(1)_\phi$.  Its couplings to
SM fermions are determined by the $U(1)_\phi$ charge assignment,
whereas those to gauge bosons are controlled by the corresponding
mixed anomalies.

In this section, we examine how the underlying $U(1)_\phi$ symmetry
constrains the visible and hidden couplings of the axion and identify
the remaining model dependence.  We first discuss the couplings of
$\phi$ to SM fermions and then the anomaly-induced couplings to visible
and dark gauge bosons.

\subsection{Couplings to Standard Model fermions}
\label{subsec:axion-fermions}

Since the Higgsino mass term involves the product $H_uH_d$, the
axion--Higgsino coupling is controlled by the sum of the
$U(1)_\phi$ charges of the two Higgs fields.  We choose the
normalization
\begin{equation}
    q_\phi(H_u)+q_\phi(H_d)=1 ,
    \label{eq:higgs-charge-sum}
\end{equation}
which fixes the normalization of the axion--Higgsino coupling used
in Sec.~\ref{sec:environment}.
The overall charge normalization is conventional, since a common
rescaling of the $U(1)_\phi$ charges can be absorbed into the axion
decay constant $f_\phi$.

After electroweak symmetry breaking, the physical axion direction
must be orthogonal to the neutral electroweak Goldstone mode eaten
by the $Z$ boson~\cite{Sun:2020iim}.  
Consequently, the physical axion couplings to SM fermions depend
only on the sum of the two Higgs charges, rather than on their
individual values.  The individual $H_u$ and $H_d$ charges need not
be specified for the fermion couplings considered below.  By
contrast, the mixed gauge-anomaly coefficients depend on the full
$U(1)_\phi$ charge assignment of the Higgs and matter fields.

For $\tan\beta\equiv v_u/v_d$, the resulting derivative couplings to
SM fermions can be written as
\begin{equation}
    {\cal L}
    =
    \frac{\partial_\mu\phi}{2f_\phi}
    \sum_f c_f\,
    \bar f\gamma^\mu\gamma_5 f ,
    \label{eq:axion-fermion-derivative}
\end{equation}
with
\begin{equation}
    c_d=c_\ell=\sin^2\beta,
    \qquad
    c_u=\cos^2\beta ,
    \label{eq:fermion-coefficients}
\end{equation}
where $u$, $d$, and $\ell$ denote up-type quarks, down-type quarks,
and charged leptons, respectively.  An overall sign of the
coefficients depends on the convention for $\phi$ and is irrelevant
for the decay rates considered below.

Using the fermion equations of motion, the interaction in
Eq.~\eqref{eq:axion-fermion-derivative} gives the fermionic decay
width of the axion, 
\begin{equation}
    \Gamma_{f\bar f}
    \equiv
    \Gamma(\phi\to f\bar f)
    =
    N_c^f\,
    \frac{c_f^2m_f^2m_\phi}{8\pi f_\phi^2}
    \sqrt{1-\frac{4m_f^2}{m_\phi^2}},
    \label{eq:phi-ff-width}
\end{equation}
where $N_c^f=3$ for quarks and $1$ for leptons. 
For quark final states, the running masses are evaluated at the scale
$m_\phi$, while the QED running of the charged-lepton masses is
negligible for the present analysis.
The up-type widths scale as $\cos^4\beta$, whereas the
down-type and charged-lepton widths scale as $\sin^4\beta$.  For
example, the relative importance of the top and bottom channels can
be written as
\begin{equation}
    \frac{\Gamma_{t\bar t}}{\Gamma_{b\bar b}}
    \simeq
    0.95
    \sqrt{
        1-\frac{4m_t^2(m_\phi)}{m_\phi^2}
    }
    \left(\frac{8}{\tan\beta}\right)^4 ,
    \label{eq:top-bottom-ratio}
\end{equation}
where the mild scale dependence of the running quark-mass ratio has
been neglected. 
The large top Yukawa coupling can make the $t\bar t$ channel important,
but its contribution decreases rapidly as $\tan\beta$ increases.
The bottom and tau couplings, by contrast, remain sizable at large
$\tan\beta$.  With the fermion couplings in Eq.~\eqref{eq:fermion-coefficients},
the $\tau^+\tau^-$ channel therefore remains an irreducible fermionic
source of high-energy solar neutrinos in this minimal realization.

\subsection{Anomalous couplings to visible and dark gauge bosons}
\label{subsec:axion-anomalies}

The axion can couple to gauge fields through mixed anomalies of
$U(1)_\phi$.  We parameterize the corresponding local interactions
as
\begin{equation}
    {\cal L}
    =
    \sum_{i=1}^{3}
    \frac{C_i\alpha_i}{8\pi}\,
    \frac{\phi}{f_\phi}\,
    F^i_{\mu\nu}\widetilde F_i^{\mu\nu}
    +
    \frac{C_D\alpha_D}{8\pi}\,
    \frac{\phi}{f_\phi}\,
    F'_{\mu\nu}\widetilde F'^{\mu\nu},
    \label{eq:axion-anomaly-couplings}
\end{equation}
where $i=1,2,3$ labels $U(1)_Y$, $SU(2)_L$, and $SU(3)_c$,
respectively.  We work in a basis in which the derivative couplings
to fermions that remain dynamical at the scale $m_\phi$ are kept
explicit, as in 
Eq.~\eqref{eq:axion-fermion-derivative}~\cite{Bauer:2020jbp}.
The coefficients $C_i$ denote the local anomaly coefficients in this
basis and depend on the gauge quantum numbers and $U(1)_\phi$ charges
of the underlying fermions.  The separation between these local
coefficients and the contributions from dynamical fermion loops is
basis dependent, whereas their combination entering physical
gauge-boson amplitudes is basis independent.
Similarly, $C_D$ denotes the local anomaly coefficient for the dark
gauge group.
We use the $SU(5)$ normalization for hypercharge, $\alpha_1=(5/3)\alpha_Y$, and identify
$\alpha_3\equiv\alpha_s$.

The physical axion couplings to visible gauge bosons also receive
loop contributions from fermions retained in the low-energy theory.
We denote by $C_i^{\rm eff}(q^2)$ the corresponding threshold-matched
coefficients entering the physical gauge-boson amplitudes, with $q$
the four-momentum carried by the axion:
\begin{equation}
    C_i^{\rm eff}(q^2)
    =
    C_i+C_i^{\rm loop}(q^2),
    \label{eq:Ceff-general}
\end{equation}
where $C_i^{\rm loop}$ denotes the contribution from dynamical
fermions. 
In the MSSM, the local QCD anomaly coefficient in the derivative
basis adopted here is $C_3=-3$ for the charge normalization in
Eq.~\eqref{eq:higgs-charge-sum}.  The derivative-coupling quark
loops contain a mass-independent anomaly contribution that cancels
this local term.  The resulting physical QCD coefficient is therefore
mass suppressed and, for the mediator masses considered here, is
well approximated by
\begin{equation}
    \left.C_3^{\rm eff}(m_\phi^2)\right|_{\rm MSSM}
    \simeq
    \cos^2\beta\,{\cal F}(\tau_t)
    +
    \sin^2\beta\,{\cal F}(\tau_b),
    \label{eq:C3eff-MSSM-leading}
\end{equation}
where $\tau_f\equiv4m_f^2/m_\phi^2$ and ${\cal F}$ denotes the
mass-dependent part of the fermion-loop function~\cite{Djouadi:2005gj}.
For $\tau_f\ll1$, it is parametrically suppressed as
${\cal F}(\tau_f)={\cal O}[\tau_f\ln^2(4/\tau_f)]$, so that the
lighter-quark contributions are negligible.

The electroweak coefficients behave differently because the
Higgsinos lie close to the kinematic threshold.  The SM-fermion
contributions are mass suppressed as in the QCD case, whereas the
near-threshold Higgsino contribution remains unsuppressed.
Although the local coefficients $C_{1,2}$ depend on the individual
$U(1)_\phi$ charge assignment, this dependence cancels against the
corresponding anomaly contributions from the dynamical fermion loops.
For the MSSM field content alone, the leading physical electroweak
coefficients in the pure-Higgsino limit, evaluated at the on-shell
mediator virtuality, are therefore
\begin{equation}
    \left.C_2^{\rm eff}(m_\phi^2)\right|_{\rm MSSM}
    \simeq
    {\cal F}(\tau_\chi),
    \qquad
    \left.C_1^{\rm eff}(m_\phi^2)\right|_{\rm MSSM}
    \simeq
    \frac{3}{5}\,{\cal F}(\tau_\chi).
    \label{eq:C12eff-MSSM}
\end{equation}
For $\Delta_\phi\ll m_\chi$, with
$m_\phi=2m_\chi-\Delta_\phi$, the loop function behaves as
\begin{equation}
    {\cal F}(\tau_\chi)
    =
    \frac{\pi^2}{4}
    -
    \pi\sqrt{\frac{\Delta_\phi}{m_\chi}}
    +
    {\cal O}\!\left(\frac{\Delta_\phi}{m_\chi}\right).
    \label{eq:higgsino-threshold-expansion}
\end{equation}
The near-threshold Higgsino loop therefore gives order-one
contributions to $C_{1,2}^{\rm eff}(m_\phi^2)$, in contrast to the
mass-suppressed QCD contribution from the lighter quarks.
For off-shell axion-mediated annihilation, the corresponding
coefficients are instead evaluated at the annihilation virtuality
$q^2=s$.

We next turn to the hidden gauge sector.  
We assume that no dark-sector fermions remain dynamical at the scale $m_\phi$, 
so that there is no separate fermion-loop contribution to the dark
gauge-boson amplitude.
The local dark anomaly interaction in
Eq.~\eqref{eq:axion-anomaly-couplings} then gives
the decay rate of the axion into a pair of dark photons as
\begin{equation}
    \Gamma_{A'A'}
    =
    \frac{(C_D\alpha_D)^2}{256\pi^3}
    \frac{m_\phi^3}{f_\phi^2}.
    \label{eq:axion-width-dark-photon}
\end{equation}
For $C_D\alpha_D$ of order unity, the dark-photon channel provides
the dominant mediator width in the parameter region of interest.
It is therefore convenient to normalize the visible partial widths
to $\Gamma_{A'A'}$.  In the decay widths below,
the effective coefficients $C_i^{\rm eff}$ are understood to be
evaluated at $q^2=m_\phi^2$.  Neglecting the electroweak
gauge-boson masses relative to $m_\phi$, we obtain 
\begin{align}
    \frac{\Gamma_{gg}}{\Gamma_{A'A'}}
    &\simeq
    8\alpha_s^2\,
    \frac{|C_3^{\rm eff}|^2}{(C_D\alpha_D)^2},
    \nonumber\\
    \frac{\Gamma_{WW}}{\Gamma_{A'A'}}
    &\simeq
    \frac{2\alpha^2}{\sin^4\theta_W}\,
    \frac{|C_2^{\rm eff}|^2}{(C_D\alpha_D)^2},
    \nonumber\\
    \frac{\Gamma_{ZZ}}{\Gamma_{A'A'}}
    &\simeq
    \alpha^2\,
    \frac{
    \left|
        \cot^2\theta_W\,C_2^{\rm eff}
        +\frac{5}{3}\tan^2\theta_W\,C_1^{\rm eff}
    \right|^2}
    {(C_D\alpha_D)^2},
    \nonumber\\
    \frac{\Gamma_{Z\gamma}}{\Gamma_{A'A'}}
    &\simeq
    2\alpha^2\,
    \frac{
    \left|
        \cot\theta_W\,C_2^{\rm eff}
        -\frac{5}{3}\tan\theta_W\,C_1^{\rm eff}
    \right|^2}
    {(C_D\alpha_D)^2},
    \nonumber\\
    \frac{\Gamma_{\gamma\gamma}}{\Gamma_{A'A'}}
    &\simeq
    \alpha^2\,
    \frac{
    \left|
        C_2^{\rm eff}
        +\frac{5}{3}C_1^{\rm eff}
    \right|^2}
    {(C_D\alpha_D)^2}.
    \label{eq:gauge-width-ratios}
\end{align}
The fermionic widths in Eq.~\eqref{eq:phi-ff-width} similarly give
\begin{equation}
    \frac{\Gamma_{f\bar f}}{\Gamma_{A'A'}}
    \simeq
    32\pi^2 N_c^f\,
    \frac{c_f^2}{(C_D\alpha_D)^2}
    \frac{m_f^2}{m_\phi^2},
    \label{eq:fermion-dark-width-ratio}
\end{equation}
for $m_f\ll m_\phi$.

For the MSSM field content alone,
$\phi$ decays into SM gauge bosons are dominated by the
electroweak final states rather than gluons.
The gluonic channel is strongly suppressed because the local
coefficient $C_3$ almost cancels the quark-loop contribution,
leaving only its mass-suppressed part.
By contrast, the near-threshold Higgsino contribution to
$C_{1,2}^{\rm eff}$ is unsuppressed, leading to the characteristic
electroweak decay pattern
\begin{equation}
    \Gamma_{WW}:
    \Gamma_{ZZ}:
    \Gamma_{\gamma\gamma}:
    \Gamma_{Z\gamma}
    \simeq
    1:0.35:0.11:0.09 ,
    \label{eq:MSSM-EW-width-ratio}
\end{equation}
where electroweak gauge-boson masses have been neglected.  The
dominant $WW$ mode satisfies approximately
\begin{equation}
    \frac{\Gamma_{WW}}{\Gamma_{A'A'}}
    \simeq
    1.6\times10^{-3}
    \left(\frac{3}{C_D\alpha_D}\right)^2 .
    \label{eq:MSSM-WW-dark-width-ratio}
\end{equation}
For $C_D\alpha_D$ of order unity, the total electroweak width
remains small compared with the dark-photon width.
The visible decay pattern can change qualitatively, however,
in the presence of additional heavy colored fermions coupled to
the axion.
Such states generate an  additional nondecoupling local
contribution to $C_3$, and a contribution of order unity can make the
gluonic mode the dominant visible decay channel.  Heavy electroweak
matter can similarly modify the relative strengths of the
electroweak channels through  additional  local
contributions to $C_{1,2}$.

 
\section{Cosmology and detection framework}
\label{sec:combined-phenomenology}

We now combine the environment-dependent annihilation derived in
Sec.~\ref{sec:environment} with the axion decay pattern discussed in
Sec.~\ref{sec:axion-couplings}.  The cosmological abundance sets the
annihilation strength required during nonthermal reannihilation, and
the environmental enhancement relates it to the corresponding rates
in the Sun and the present Galactic halo.  The axion decay pattern
then determines how these low-velocity annihilations manifest
themselves in solar-neutrino and Galactic gamma-ray searches.  We
first establish the connection to the nonthermal Higgsino abundance
and then examine the complementary detection signatures.

\subsection{Nonthermal Higgsino reannihilation}
\label{subsec:nonthermal-higgsino}

We consider a cosmological history in which the Higgsino abundance
undergoes reannihilation at a temperature $T_{\rm ann}$.  The
population entering this epoch can contain both thermally produced
Higgsinos and Higgsinos injected by late decays.  If the total
abundance is sufficiently large, efficient reannihilation renders
the final abundance largely insensitive to the preceding production
history~\cite{Moroi:1999zb,Endo:2006ix,Fukuda:2024dvi}.  It is then controlled primarily by the annihilation rate
at $T_{\rm ann}$.

Since $R_{\rm cos}$ measures the enhancement of the total
cosmological annihilation rate over the coannihilation-inclusive
electroweak rate, efficient reannihilation directly relates
$T_{\rm ann}$, $R_{\rm cos}$, and the Higgsino fraction $\xi_\chi$.
In the nearly degenerate pure-Higgsino limit, the cosmological
electroweak annihilation rate scales as $m_\chi^{-2}$.  Together
with the standard reannihilation relation,
$\Omega_\chi \propto
m_\chi\bigl(T_{\rm ann}\langle\sigma v\rangle_{\rm cos}\bigr)^{-1}$,
this implies that the required enhancement at fixed $T_{\rm ann}$
and $\xi_\chi$ grows approximately as $m_\chi^3$.  For the range of
$T_{\rm ann}$ considered here, the relic-abundance relation is
approximately
\begin{equation}
    R_{\rm cos}
    \simeq
    93
    \left(
        \frac{m_\chi}{500~{\rm GeV}}
    \right)^3
    \left(
        \frac{1.5~{\rm GeV}}{T_{\rm ann}}
    \right)
    \left(
        \frac{0.03}{\xi_\chi}
    \right),
    \label{eq:Rcos-Tann-xi}
\end{equation} 
where the mild residual dependence on the effective relativistic
degrees of freedom is not kept explicitly.  We therefore
parameterize the cosmological history in terms of $T_{\rm ann}$ and
$\xi_\chi$, with $R_{\rm cos}$ determined by the relic-abundance
requirement.

The relic-abundance condition also fixes the axion decay constant
required for reannihilation through
\begin{equation}
    \left\langle \sigma v \right\rangle_{\phi,\rm eff}^{\rm cos}
    =
    (R_{\rm cos}-1)
    \left\langle \sigma v \right\rangle_{\rm EW,eff}^{\rm cos},
    \label{eq:fphi-relic-condition}
\end{equation}
where the axion-mediated rate includes the thermal average and the
Higgsino population factors discussed in
Sec.~\ref{subsec:environmental-enhancement}.
As in Sec.~\ref{subsec:environmental-enhancement}, we neglect the
subleading interference correction to the total cosmological
annihilation rate.

In the nearly degenerate Higgsino and narrow-width limits, the
environmental enhancement derived there gives the useful estimate
\begin{align}
    f_\phi \simeq{}&
    1.1~{\rm TeV}
    \left(\frac{C_D\alpha_D}{3}\right)^{1/2}
    \left(\frac{\xi_\chi}{0.03}\right)^{1/4}
    \left(\frac{m_\chi}{500~{\rm GeV}}\right)^{1/4}
    \left(\frac{1.5~{\rm GeV}}{T_{\rm ann}}\right)^{1/4}
    \nonumber\\
    &\times
    \left[
        \frac{
            (\Delta_\phi/T_{\rm ann})^2
            +3.3(\Delta_\phi/T_{\rm ann})
        }{6}
    \right]^{-1/4}.
    \label{eq:fphi-relic-estimate}
\end{align}
Thus the relic-abundance condition fixes $f_\phi$ near the TeV scale
in the parameter region of interest.  The corresponding Higgsino
coupling $m_\chi/f_\phi$ remains perturbative, while the mediator mass
lies below the expected strong-coupling scale of order $4\pi f_\phi$.
Equation~\eqref{eq:fphi-relic-estimate} neglects finite-width effects.
At the reference point, $\Gamma_\phi/\Delta_\phi\simeq 0.5$, but the
width enters the near-threshold propagator through
$\Delta_\phi^2+\Gamma_\phi^2/4$ in the zero-velocity limit.
The resulting correction is therefore controlled by
$\Gamma_\phi^2/(4\Delta_\phi^2)$ and modifies the environmental
enhancement only at the few-percent level.

We next relate the cosmological annihilation enhancement to that in
low-velocity environments.  Using Eq.~\eqref{eq:environment-final},
$R_\odot$ and $R_{\rm halo}$ are determined, to the accuracy
discussed in Sec.~\ref{subsec:environmental-enhancement}, by the
value of $R_{\rm cos}$ required by the relic abundance.  An
axion-mediated contribution that is moderate during cosmological
reannihilation can therefore become dominant in both low-velocity
environments, with complementary implications for solar-neutrino
and Galactic-halo searches.


\subsection{Solar capture and neutrino signal}
\label{subsec:solar-neutrino}

The quasi-Dirac Higgsino can be captured in the Sun through the
off-diagonal $Z$ interaction responsible for inelastic
Higgsino--nucleus scattering.  Solar capture requires not only that
the incoming Higgsino have sufficient kinetic energy to up-scatter
into the heavier neutral state, but also that the inelastic
scattering remove enough energy for the Higgsino to become
gravitationally bound.  The resulting capture rate $C_\odot$
therefore depends sensitively on $m_\chi$, $\Delta m_0$, and the
solar nuclear composition~\cite{Menon:2009qj}.

Assuming that the captured Higgsinos thermalize in the solar core,
the annihilation rate can be written as
\begin{equation}
    \Gamma_{\rm ann}^{\odot}
    =
    \frac{C_\odot}{2}
    \tanh^2\left(
        \frac{t_\odot}{\tau_{\rm eq}}
    \right),
    \label{eq:solar-equilibration}
\end{equation}
where $t_\odot$ is the solar age and
$\tau_{\rm eq}=(C_\odot C_A)^{-1/2}$ is the equilibration time.
Here $C_A$ denotes the annihilation coefficient of the captured
Higgsino population.  In the capture--annihilation equilibrium
limit, $t_\odot\gg\tau_{\rm eq}$, Eq.~\eqref{eq:solar-equilibration}
reduces to
\begin{equation}
    \Gamma_{\rm ann}^{\odot}
    \simeq
    \frac{1}{2}C_\odot .
    \label{eq:solar-equilibrium}
\end{equation}
The capture rate $C_\odot$ scales linearly with the Higgsino fraction
$\xi_\chi$.

In generic pseudo-Dirac dark-matter models, two-state kinetics can
strongly suppress annihilation when it requires a thermally depleted
excited state~\cite{DiMauro:2026dqp}.  This suppression does
not occur for the Higgsino considered here.  The lightest neutral
state annihilates directly through the standard electroweak channels,
while the axion-mediated interaction is likewise diagonal in the
neutral-Higgsino mass basis.
For representative parameters in the efficient-capture regime, we
find that the standard electroweak annihilation rate is sufficient
to establish capture--annihilation equilibrium down to Higgsino
fractions of order $\xi_\chi\sim10^{-6}$~\cite{Pospelov:2026ewn},
although the precise threshold depends on $m_\chi$ and $\Delta m_0$
through the capture rate.
The axion-mediated channel further shortens the equilibration time
by enhancing the low-velocity annihilation rate.

Once capture--annihilation equilibrium is established, an enhancement
of the low-velocity annihilation cross section does not increase the
total annihilation rate.  
Instead, it changes the relative fractions
of annihilations into the available final states and hence the
resulting neutrino signal.  Interference must be included whenever
the standard Higgsino and axion-mediated amplitudes contribute to the
same final state.  For a final state $X$, we write the physical
annihilation rate as
\begin{equation}
    \langle\sigma v\rangle_X
    =
    \langle\sigma v\rangle_{{\rm EW},X}
    +
    (1+\kappa_X)
    \langle\sigma v\rangle_{\phi,X},
    \label{eq:solar-interference}
\end{equation}
where
\begin{equation}
    \kappa_X
    \equiv
    \frac{
        \langle\sigma v\rangle_{{\rm int},X}
    }{
        \langle\sigma v\rangle_{\phi,X}
    } .
    \label{eq:kappa-interference}
\end{equation}
parameterizes the interference between the standard Higgsino and
axion-mediated amplitudes.  Negative values of $\kappa_X$ correspond
to destructive interference, whereas positive values correspond to
constructive interference.  For sufficiently strong destructive
interference, the combined axion-mediated and interference
contribution can become negative, while the physical annihilation
rate in Eq.~\eqref{eq:solar-interference} remains nonnegative.
Sizable interference is relevant primarily for the $WW$ and $ZZ$
final states, where the standard Higgsino and axion-mediated
amplitudes populate the same gauge-boson helicity configuration in
the low-velocity limit.  For the other channels, there is no
comparably large standard Higgsino amplitude contributing to the
same low-velocity final-state configuration, and the corresponding
interference is subleading.

In capture--annihilation equilibrium, the total annihilation rate is
fixed by the capture rate, while the relative contribution of each
final state is determined by its physical annihilation rate
$\langle\sigma v\rangle_X$.  The high-energy solar-neutrino signal
relative to the standard Higgsino reference is therefore
\begin{equation}
    F_{\nu,\odot}
    \equiv
    \frac{\Phi_\nu}{\Phi_\nu^{\rm std}}
    =
    \frac{
        \sum_X
        \epsilon_X\,\langle\sigma v\rangle_X
    }{
        \sum_X
        \langle\sigma v\rangle_X
    } .
    \label{eq:solar-neutrino-suppression}
\end{equation}
We denote by $\epsilon_X$ the IceCube solar-neutrino signal yield
per annihilation into the final state $X$, normalized to the
annihilation-weighted yield of the standard Higgsino in the Sun.
The weight $\epsilon_X$ accounts for the neutrino spectrum produced
by the final state, propagation through the Sun, and the
energy-dependent detector response.  With this convention,
$F_{\nu,\odot}=1$ for standard Higgsino annihilation.  For the
stable dark-photon final state, $\epsilon_{A'A'}=0$.

Using Eq.~\eqref{eq:solar-interference} and the definition of
$R_\odot$ in Eq.~\eqref{eq:Rsun-general}, the normalized neutrino
signal can be decomposed as
\begin{equation}
\begin{split}
    F_{\nu,\odot}
    ={}&
    \frac{1}{R_\odot}
    +
    \frac{1}{R_\odot}
    \frac{
        \sum_X
        \epsilon_X(1+\kappa_X)
        \langle\sigma v\rangle_{\phi,X}
    }{
        \langle\sigma v\rangle_{\rm EW}
    } .
    \label{eq:solar-neutrino-residual}
\end{split}
\end{equation}
The first term represents the residual standard-Higgsino
contribution, suppressed by $1/R_\odot$, while the second term
contains the neutrino-weighted axion-mediated contribution including
its interference with the standard Higgsino amplitudes.  Individual
channel contributions to the second term can be negative for
sufficiently strong destructive interference, while the full
neutrino signal remains nonnegative.

The electroweak gauge-boson, $t\bar t$, and $\tau^+\tau^-$ channels
produce hard neutrino spectra whose production and propagation in
the Sun are well studied~\cite{Baratella:2013fya}.  Their
contributions to the IceCube solar-neutrino signal are of the same
order as that from standard Higgsino annihilation, but differ by
order-one factors among the individual channels.  
In the minimal realization considered here, the
$\tau^+\tau^-$ channel is particularly important when the
electroweak gauge-boson modes are suppressed, since its coupling
remains sizable at large $\tan\beta$.

At $m_\chi=500~{\rm GeV}$, the published ten-year IceCube limits
give a relative signal yield of approximately $1.2$ for
$\tau^+\tau^-$ compared with $W^+W^-$~\cite{IceCube:2025fcu}.
For channels without a dedicated IceCube limit, we fold the
PPPC~4~DM$\nu$ spectra with the public IceCube effective-area
information~\cite{Baratella:2013fya,IceCube:2025fcu}.
This gives a detector-response-level $ZZ$ yield approximately
$1.2$ times that of $W^+W^-$.  Normalizing these yields to the
annihilation-weighted yield of the standard Higgsino gives
$\epsilon_{WW}\simeq0.9$, $\epsilon_{ZZ}\simeq1.1$, and
$\epsilon_\tau\simeq1.1$.  The $ZZ$ value should be understood as
a response-level estimate rather than a result of the full IceCube
likelihood.

By contrast, the neutrino spectra from the $b\bar b$ and gluonic
channels are strongly softened by hadronization and energy loss in
the solar medium.  Existing studies indicate that their
contributions to the high-energy IceCube signal are strongly
suppressed relative to the hard channels, with representative
suppressions of order $\epsilon_b\sim10^{-2}$ and
$\epsilon_{gg}\sim10^{-3}$~\cite{Ibarra:2013eba,IceCube:2016yoy}.
The precise values depend on the Higgsino mass and detector response,
but the neutrino yields follow the schematic hierarchy
$\epsilon_{gg}\ll\epsilon_b\ll\epsilon_{\rm hard}\sim1$.

The relative importance of these channels is determined by the axion
couplings discussed in Sec.~\ref{subsec:axion-anomalies}.  For the
MSSM field content alone, the near-threshold Higgsino contribution
gives unsuppressed electroweak gauge-boson modes, with $WW$ providing
the largest partial width.
The fermionic channels are instead controlled by the $U(1)_\phi$
charges discussed in Sec.~\ref{subsec:axion-fermions}.
With the fermion couplings in Eq.~\eqref{eq:fermion-coefficients},
the $t\bar t$ contribution decreases rapidly with increasing
$\tan\beta$, while in this minimal realization the $\tau^+\tau^-$
mode remains an irreducible source of hard neutrinos at large
$\tan\beta$.
The gluonic mode is strongly suppressed in the MSSM, but can become
important if additional heavy colored matter generates a sizable
local QCD anomaly contribution.
Additional heavy electroweak matter can likewise modify the
magnitude and relative sign of the axion-mediated $WW$ and $ZZ$
amplitudes through local contributions to $C_{1,2}^{\rm eff}$.
The solar-neutrino constraint must therefore be evaluated using the
neutrino-weighted physical annihilation rates in
Eq.~\eqref{eq:solar-neutrino-suppression}, rather than from the
axion decay pattern alone.

For given $T_{\rm ann}$ and $\xi_\chi$, the relic-abundance
requirement fixes $R_{\rm cos}$ and hence, to the accuracy discussed
in Sec.~\ref{subsec:environmental-enhancement}, $R_\odot$.  The
solar-neutrino signal then follows from
Eq.~\eqref{eq:solar-neutrino-suppression}, including the
channel-dependent neutrino response and interference effects.
Capture--annihilation equilibrium fixes the total annihilation rate,
while the observable neutrino flux is determined by the physical
annihilation rates into the individual final states.


\subsection{Gamma rays from Galactic-halo annihilation}
\label{subsec:halo-gamma}

As shown in Sec.~\ref{sec:environment}, the present Galactic halo
probes essentially the same low-velocity Higgsino annihilation rate
as the Sun, so that $R_{\rm halo}\simeq R_\odot$.  Unlike in the Sun,
however, the Galactic-halo annihilation rate depends directly on the
low-velocity annihilation cross section.  Assuming
that the local Higgsino fraction follows its cosmological abundance
fraction, the halo signal scales as $\xi_\chi^2$.  For a final state
$X$, we therefore define the effective annihilation rate
\begin{equation}
    \langle\sigma v\rangle_X^{\rm eff}
    \equiv
    \xi_\chi^2
    \left[
        \langle\sigma v\rangle_{{\rm EW},X}^{\rm halo}
        +
        (1+\kappa_X)
        \langle\sigma v\rangle_{\phi,X}^{\rm halo}
    \right],
    \label{eq:halo-effective-rate}
\end{equation}
where $\kappa_X$ is the interference parameter defined in
Eq.~\eqref{eq:kappa-interference}.  As in the solar environment,
interference is relevant primarily for the $WW$ and $ZZ$ final
states and is subleading for the other channels.
For the gamma-ray-producing final states considered below,
$\langle\sigma v\rangle_X^{\rm eff}$ can be compared directly with
indirect-detection limits obtained under the assumption that the
annihilating particle constitutes the full dark-matter density.

The axion-mediated contribution to the halo
annihilation rate has a simple near-threshold scaling.  In the
narrow-width regime $\Gamma_\phi\ll\Delta_\phi$, the zero-velocity
rate into an axion-mediated final state $X$ scales parametrically as
$\xi_\chi^2\langle\sigma v\rangle_{\phi,X}^{\rm halo}
\propto\xi_\chi^2\Gamma_X/(f_\phi\Delta_\phi)^2$, up to factors
that vary slowly near threshold.  For fixed dimensionless
couplings, $\Gamma_X\propto f_\phi^{-2}$, so that
the axion-mediated contributions to the individual
final states share the common dependence
\begin{equation}
    \xi_\chi^2
    \langle\sigma v\rangle_{\phi,X}^{\rm halo}
    \propto
    \Big(
        \frac{\xi_\chi}{f_\phi^2\Delta_\phi}
    \Big)^2,
    \label{eq:halo-rate-scaling}
\end{equation}
with the remaining channel dependence determined by the
corresponding fermion or anomaly couplings and phase-space factors.
For the $WW$ and $ZZ$ final states, this scaling applies only to the
axion-mediated contribution, while the physical rates also include
the standard Higgsino contribution and its interference with the
axion-mediated amplitude.

The visible axion-mediated channels produce both continuum and
monochromatic gamma-ray signals.  Continuum emission arises from
$gg$, $b\bar b$, $t\bar t$, $W^+W^-$, $ZZ$, and
$\tau^+\tau^-$ final states through subsequent showering and decay.
The corresponding photon spectra for standard two-body final states
are well studied; representative results can be obtained, for example,
from PPPC 4 DM ID~\cite{Cirelli:2010xx}.

For the MSSM field content alone, the gluonic channel is strongly
suppressed because $C_3^{\rm eff}$ is generated by mass-suppressed
quark loops.  The visible gauge-boson partial widths are instead
dominated by the electroweak channels generated by the near-threshold
Higgsino loop, with $W^+W^-$ providing the largest partial width.
A sizable gluonic continuum can nevertheless arise if additional
colored fermions much heavier than the axion generate
an order-one local QCD anomaly contribution.
This possibility is particularly interesting because the gluonic
channel produces an appreciable continuum gamma-ray spectrum while
giving only a small high-energy solar-neutrino yield.

In addition to the continuum emission, the electroweak gauge-boson
couplings produce monochromatic gamma-ray lines through the
$\gamma\gamma$ and $Z\gamma$ channels.
For nonrelativistic halo Higgsinos, the corresponding photon energies
are
\begin{equation}
    E_\gamma^{\gamma\gamma}
    \simeq m_\chi,
    \qquad
    E_\gamma^{Z\gamma}
    \simeq
    m_\chi
    \left(
        1-\frac{m_Z^2}{4m_\chi^2}
    \right).
    \label{eq:halo-line-energies}
\end{equation}
For the MSSM field content alone, the same near-threshold Higgsino
loop determines the $WW$, $ZZ$, $\gamma\gamma$, and $Z\gamma$
channels.
The gamma-ray line signals are therefore correlated
with the electroweak continuum through the same matched electroweak
coefficients, although the relation between their observable rates
is not simply given by the mediator decay pattern.
Additional heavy electroweak matter can modify the relative strengths
of these channels through local contributions to $C_{1,2}$.

The dominant dark-photon channel has qualitatively
different consequences in the two low-velocity environments.
In the Sun, increasing the hidden branching fraction reduces the
fraction of annihilations producing high-energy neutrinos, whereas
the Galactic signal from a final state $X$ is determined by its
partial annihilation rate.

Once the cosmological abundance and environmental enhancement are
specified, the total axion-mediated annihilation rate in the halo is
fixed, while its contribution to a visible final state $X$ scales
with ${\rm Br}(\phi\to X)$.  Increasing $C_D\alpha_D$ therefore
shifts the mediator decays toward dark photons and suppresses the
axion-mediated visible channels.  The compact environmental relation
used here assumes the narrow-width regime
$\Gamma_\phi\ll\Delta_\phi$.  As the mediator width approaches the
threshold detuning, the width term regulates the near-threshold
propagator and weakens the velocity-dependent environmental
enhancement, in which case the full finite-width thermal average
must be used.

The Galactic-halo phenomenology can therefore be organized in terms
of parameters with distinct physical roles.  For a specified
parameter point, the relic-abundance condition fixes $R_{\rm cos}$
and the required $f_\phi$, while the environmental enhancement
determines $R_{\rm halo}$.  The ratio $\Delta_\phi/T_{\rm ann}$
controls the velocity-dependent part of this enhancement.  The
matched coefficients $C_i^{\rm eff}$ and $C_D\alpha_D$, together
with $\tan\beta$, determine the relative strengths of the visible
and hidden axion-mediated channels.  Together with the standard
Higgsino contribution, these determine the physical continuum and
line gamma-ray signals.  Finally, $\Gamma_\phi/\Delta_\phi$
controls the validity of the narrow-width description and the onset
of finite-width suppression of the velocity-dependent enhancement.


\section{Phenomenological implications}
\label{sec:phenomenological-implications}

We now apply the framework developed in the previous sections to the
observable consequences of environment-dependent annihilation.
Since solar-neutrino searches impose a particularly stringent
constraint on quasi-Dirac Higgsinos in the efficient-capture regime,
we begin by identifying the decay channels that dominate the
solar-neutrino signal and determining the conditions required to
satisfy the IceCube bound.  We next examine the associated continuum
and line gamma-ray signals from the Galactic halo, assessing their
potential observability as well as the constraints imposed by
existing indirect searches.
Combining the solar-neutrino and gamma-ray constraints, we identify
the conditions under which a sizable Higgsino fraction can be
retained, before examining the implications for terrestrial
inelastic searches.


\subsection{Solar-neutrino constraints}
\label{subsec:solar-neutrino-constraints}

We begin with the solar-neutrino constraint, which provides a
particularly stringent test of quasi-Dirac Higgsinos in the
efficient-capture regime.  To compare the predicted signal directly
with the IceCube limit, we define
\begin{equation}
    S_\odot
    \equiv
    \xi_\chi F_{\nu,\odot},
    \label{eq:solar-signal-strength}
\end{equation}
where $F_{\nu,\odot}$ is the normalized solar-neutrino signal defined
in Eq.~\eqref{eq:solar-neutrino-suppression}.  Here $S_\odot$ is the
high-energy solar-neutrino signal normalized to that of a standard
Higgsino with the same $m_\chi$ and $\Delta m_0$ constituting the
full dark-matter abundance.  The factor $\xi_\chi$ accounts for the
reduced capture rate of a subdominant Higgsino component, while
$F_{\nu,\odot}$ accounts for the modification of the neutrino yield
by the annihilation pattern after capture.  Thus $S_\odot=1$
corresponds to the full-density standard-Higgsino reference.  The
IceCube constraint can be written as
\begin{equation}
    S_\odot
    <
    s_\odot^{\rm lim}(m_\chi,\Delta m_0),
    \label{eq:IceCube-general-constraint}
\end{equation}
where $s_\odot^{\rm lim}$ is the experimental upper limit normalized
to the same standard-Higgsino reference.  For the reference point
$m_\chi=500~{\rm GeV}$ and $\Delta m_0=320~{\rm keV}$, the IceCube
limit corresponds to
\begin{equation}
    s_\odot^{\rm lim}
    \simeq
    8\times10^{-5},
    \label{eq:solar-limit-reference}
\end{equation}
with a representative range of
$(7\text{--}10)\times10^{-5}$~\cite{IceCube:2025fcu,Nguyen:2026lui}.
The limit depends on $\Delta m_0$ through the inelastic solar-capture
rate and should be evaluated at the corresponding splitting when
considering other values of $\Delta m_0$.

To identify the origin of the solar-neutrino signal,
it is useful to decompose Eq.~\eqref{eq:solar-neutrino-residual}
into the residual standard-Higgsino contribution and the
channel-dependent contributions induced by the axion interaction.
Using Eq.~\eqref{eq:solar-signal-strength}, we write
\begin{equation}
    S_\odot
    =
    S_{\odot,{\rm std}}
    +
    \sum_X S_{\odot,X},
    \label{eq:solar-signal-decomposition}
\end{equation}
where
\begin{equation}
    S_{\odot,{\rm std}}
    \equiv
    \frac{\xi_\chi}{R_\odot},
    \qquad
    S_{\odot,X}
    \equiv
    \frac{\xi_\chi}{R_\odot}\,
    \epsilon_X(1+\kappa_X)
    \frac{
        \langle\sigma v\rangle_{\phi,X}
    }{
        \langle\sigma v\rangle_{\rm EW}
    } .
    \label{eq:solar-signal-components}
\end{equation}
The first term is the residual standard-Higgsino
signal and is suppressed by $1/R_\odot$.  The quantity
$S_{\odot,X}$ instead measures the change in the neutrino signal
associated with the axion-mediated amplitude in the final state
$X$, including its interference with the standard Higgsino
amplitude and weighted by the corresponding neutrino response
$\epsilon_X$.  It should therefore not be interpreted as a
positive-definite physical channel rate and can be negative for
sufficiently strong destructive interference.
 
The strength of the IceCube constraint is particularly clear from
the normalization of $S_\odot$.  A standard Higgsino constituting
the full dark-matter abundance has $S_\odot=1$, whereas
solar-neutrino searches in the efficient-capture regime probe
$S_\odot$ at the $10^{-4}$ level and below.  Since
$F_{\nu,\odot}=1$ for standard Higgsino annihilation, satisfying
such a bound through abundance dilution alone would require
$\xi_\chi$ to be comparably small. 
The environment-dependent annihilation dynamics can instead reduce
$F_{\nu,\odot}$ by changing the relative rates into the different
final states, allowing a substantially larger Higgsino fraction
while suppressing the observable solar-neutrino signal.

  
We next quantify the interference in the electroweak gauge-boson
channels.  At zero velocity, two identical Majorana Higgsinos form
a spin-singlet $S$-wave initial state~\cite{Jungman:1995df,
Nihei:2002ij,Beneke:2012tg}.  The corresponding annihilation
amplitude lies in the CP-odd $J=0$ channel and populates the same
transverse two-vector state as the axion operator
$\phi W_{\mu\nu}\widetilde W^{\mu\nu}$.  The standard Higgsino and
axion-mediated $WW$ amplitudes therefore have complete helicity
overlap in this limit.  With the normalization of
Eq.~\eqref{eq:kappa-interference}, this gives
\begin{equation}
    \kappa_{WW}
    =
    2\eta_{WW}
    \left(
        \frac{
            \langle\sigma v\rangle_{{\rm EW},WW}
        }{
            \langle\sigma v\rangle_{\phi,WW}
        }
    \right)^{1/2}
    |\cos\delta_{WW}|,
    \label{eq:kappa-WW-general}
\end{equation}
where $\eta_{WW}=\pm1$ specifies the relative sign of the two
amplitudes and $\delta_{WW}$ denotes the residual phase associated
with the mediator width.  Evaluating the chargino-exchange and
Higgsino-loop amplitudes in a common convention gives
$\eta_{WW}=-1$ for the subthreshold mediator, corresponding to
destructive interference.

The Higgsino-loop coefficient entering the axion-mediated
annihilation amplitude must be evaluated at the annihilation
virtuality $q^2=s$, rather than at the mediator pole
$q^2=m_\phi^2$~\cite{Bauer:2020jbp,Bonilla:2021ufe}. 
For the MSSM field content and the charge normalization
$q_\phi(H_u)+q_\phi(H_d)=1$ adopted in
Eq.~\eqref{eq:higgs-charge-sum}, the static-limit coefficient is
$C_2^{\rm eff}(s=4m_\chi^2)=\pi^2/4$.  In the narrow-width limit,
this gives
\begin{equation}
    \left.\kappa_{WW}\right|_{\rm MSSM}
    \simeq
-32\frac{f_\phi^2 \Delta_\phi}{m_\chi^3}
    \simeq
    -0.62
    \left(
        \frac{f_\phi}{1.1~{\rm TeV}}
    \right)^2
    \left(
        \frac{\Delta_\phi}{2~{\rm GeV}}
    \right)
    \left(
        \frac{500~{\rm GeV}}{m_\chi}
    \right)^3 .
    \label{eq:kappa-WW-MSSM}
\end{equation}
Finite-width effects modify this relation through the ratio
$\Gamma_\phi/\Delta_\phi$ and are subleading in the parameter region
considered here.

The $ZZ$ channel has the same destructive sign in the pure-Higgsino
limit.  Before electroweak mixing, the initial two-Higgsino state
contains no isospin-two component, whereas the symmetric $J=0$ state of the two SU(2)$_L$ gauge bosons
decomposes into an 
SU(2)$_L$ singlet and a
five-plet.  The five-plet amplitude is therefore absent, leaving the
same singlet structure for the standard and axion-mediated
amplitudes.  
Including hypercharge with the MSSM Higgsino couplings preserves
the destructive relative sign, so that
$\eta_{ZZ}=\eta_{WW}=-1$.
In the degenerate pure-Higgsino limit, neglecting the electroweak
gauge-boson masses, the zero-velocity annihilation rates satisfy
\[
    \frac{\langle\sigma v\rangle_{{\rm EW},ZZ}}
         {\langle\sigma v\rangle_{{\rm EW},WW}}
    =
    \frac{1}{2\cos^4\theta_W},
    \qquad
    \frac{\langle\sigma v\rangle_{\phi,ZZ}}
         {\langle\sigma v\rangle_{\phi,WW}}
    =
    \frac{(\cos^4\theta_W+\sin^4\theta_W)^2}
         {2\cos^4\theta_W}.
\]
The first ratio follows from Eq.~(2.20), while the second follows
from Eq.~(3.12) using
$C_1^{\rm eff}(s)=\frac{3}{5}C_2^{\rm eff}(s)$.
Since the two channels have the same interference sign and share
the mediator phase, applying Eq.~(5.6) to both channels gives
\begin{equation}
\begin{aligned}
    \left.\kappa_{ZZ}\right|_{\rm MSSM}
    &=
    \left[
        \frac{
            \langle\sigma v\rangle_{{\rm EW},ZZ}
            /
            \langle\sigma v\rangle_{{\rm EW},WW}
        }{
            \langle\sigma v\rangle_{\phi,ZZ}
            /
            \langle\sigma v\rangle_{\phi,WW}
        }
    \right]^{1/2}
    \left.\kappa_{WW}\right|_{\rm MSSM}
    \\[2pt]
    &=
    \frac{\left.\kappa_{WW}\right|_{\rm MSSM}}
         {\cos^4\theta_W+\sin^4\theta_W}
    \simeq
    1.55\,\left.\kappa_{WW}\right|_{\rm MSSM}.
     \label{eq:ZZ-WW-interference-relation}
\end{aligned}
\end{equation}
The interference therefore substantially modifies the visible
electroweak channels, with an especially strong cancellation in the
$ZZ$ channel for the reference parameters considered below.  More
generally, sufficiently strong destructive interference can make
the combined axion-mediated and interference contribution negative.

More generally, additional heavy electroweak matter modifies the
axion-mediated amplitudes through the matched coefficients
$C_{1,2}^{\rm eff}(s)$.  Keeping the standard Higgsino amplitudes
fixed, the interference parameters satisfy
\begin{equation}
\begin{split}
    \kappa_{WW}
    &=
    \frac{
        \left.C_{2}^{\rm eff}(s)\right|_{\rm MSSM}
    }{
        C_2^{\rm eff}(s)
    }
    \left.\kappa_{WW}\right|_{\rm MSSM},
    \\[2mm]
    \kappa_{ZZ}
    &=
    \frac{
        C_2^{\rm eff}(s)
    }{
        \cos^4\theta_W\,C_2^{\rm eff}(s)
        +\frac{5}{3}\sin^4\theta_W\,C_1^{\rm eff}(s)
    }
    \kappa_{WW}.
    \label{eq:general-EW-interference}
\end{split}
\end{equation}
Additional electroweak contributions can therefore modify both the
magnitude and the sign of the interference.  For the MSSM Higgsino
coefficients, these relations reduce to
Eq.~\eqref{eq:ZZ-WW-interference-relation}.

We first consider the residual standard-Higgsino contribution, which
is independent of the mediator decay pattern.  In the regime
$R_{\rm cos}\gg1$, combining the relic-abundance relation with the
environmental enhancement derived in Sec.~2 gives
\begin{equation}
    S_{\odot,\rm std}
    \simeq
    7.3\times10^{-6}
    \left(\frac{\xi_\chi}{0.03}\right)^2
    \left(\frac{500~{\rm GeV}}{m_\chi}\right)^3
    \left(\frac{T_{\rm ann}}{1.5~{\rm GeV}}\right)
    \left(
        \frac{
            1+3.3\,T_{\rm ann}/\Delta_\phi
        }{
            3.5
        }
    \right)^{-1}.
    \label{eq:solar-standard-estimate}
\end{equation}
The numerical coefficient corresponds to the nearly degenerate
Higgsino limit, $K_0\simeq12.4$, with the reference normalization
$T_{\rm ann}=1.5~{\rm GeV}$ and $\Delta_\phi=2~{\rm GeV}$.
Away from this limit, $K_0$ and the effective electroweak
coannihilation rate should be evaluated with the appropriate
charged-Higgsino Boltzmann factor.

We now turn to the contributions from the individual visible
channels.  In the dark-width-dominated regime,
$\Gamma_\phi\simeq\Gamma_{A'A'}$, the axion-mediated rates into
visible final states are controlled by the corresponding partial
widths relative to $\Gamma_{A'A'}$.  Since all mediator partial
widths share the same overall $f_\phi^{-2}$ dependence, these rate
ratios are independent of $f_\phi$.  The $WW$ and $ZZ$
contributions to the solar-neutrino signal additionally depend on
their interference with the standard Higgsino amplitudes.

For the electroweak channels, the matched coefficients entering
the annihilation amplitudes are evaluated at the annihilation
virtuality $q^2=s$.  For the MSSM field content, the zero-velocity
limit relevant for solar annihilation, $s\simeq4m_\chi^2$, gives
$C_2^{\rm eff}=\pi^2/4$ and
$C_1^{\rm eff}=3\pi^2/20$.
Interference in the remaining channels is subleading and will be
neglected in the estimates below.  The leading contributions from
the hard-neutrino channels are then given by
\begin{align}
S_{\odot,WW}
&\simeq
4.5\times10^{-5}\,
\epsilon_{WW}\,(1+\kappa_{WW})
\left(\frac{\xi_\chi}{0.03}\right)
\left(\frac{C_2^{\rm eff}}{2.47}\right)^2
\left(\frac{3}{C_D\alpha_D}\right)^2 ,
\nonumber\\
S_{\odot,ZZ}
&\simeq
1.6\times10^{-5}\,
\epsilon_{ZZ}\,(1+\kappa_{ZZ})
\left(\frac{\xi_\chi}{0.03}\right)
\left(
    \frac{C_2^{\rm eff}+0.15\,C_1^{\rm eff}}{2.69}
\right)^2
\left(\frac{3}{C_D\alpha_D}\right)^2 ,
\nonumber\\
S_{\odot,Z\gamma}
&\simeq
3.8\times10^{-6}\,
\epsilon_{Z\gamma}
\left(\frac{\xi_\chi}{0.03}\right)
\left(
    \frac{C_2^{\rm eff}-0.5\,C_1^{\rm eff}}{1.73}
\right)^2
\left(\frac{3}{C_D\alpha_D}\right)^2 ,
\nonumber\\
S_{\odot,t\bar t}
&\simeq
1.5\times10^{-5}\,
\epsilon_t
\left(\frac{\xi_\chi}{0.03}\right)
\left(\frac{500~{\rm GeV}}{m_\chi}\right)^2
\left(\frac{8}{\tan\beta}\right)^4
\left(\frac{3}{C_D\alpha_D}\right)^2 ,
\nonumber\\
S_{\odot,\tau\tau}
&\simeq
3.3\times10^{-6}\,
\epsilon_\tau
\left(\frac{\xi_\chi}{0.03}\right)
\left(\frac{500~{\rm GeV}}{m_\chi}\right)^2
\left(\frac{3}{C_D\alpha_D}\right)^2 .
\label{eq:solar-hard-channel-estimates}
\end{align}
For the fermionic channels, the additional mass dependence follows from
$\Gamma_{f\bar f}/\Gamma_{A'A'}\propto m_f^2/m_\phi^2$, with
$m_\phi\simeq 2m_\chi$ in the near-threshold limit.
For the MSSM coefficients and the normalization in
Eq.~\eqref{eq:kappa-WW-MSSM}, the interference factors are
$1+\kappa_{WW}\simeq0.38$ and
$1+\kappa_{ZZ}\simeq0.04$.  The destructive interference therefore
reduces the $WW$ contribution to about $40\%$ of its incoherent
value and nearly cancels the axion-induced contribution in the
$ZZ$ channel.
Since the $ZZ$ channel lies close to a cancellation point,
its residual axion-induced contribution is sensitive to small variations
in the matched electroweak coefficients as well as to electroweak
Sommerfeld corrections to the standard Higgsino amplitude.
It should therefore not be regarded as a precision prediction. 
More generally, the combined axion-mediated and interference
contribution can be negative, corresponding to a reduction of the
physical channel rate relative to its standard-Higgsino value.
The physical rate itself remains positive after the standard
contribution is included.

The $b\bar b$ and gluonic channels are less important for the
high-energy solar-neutrino signal because of their much softer
neutrino spectra.  The bottom-quark contribution can be estimated as
\begin{equation}
    S_{\odot,b\bar b}
    \simeq
    1.8\times10^{-7}
    \left(\frac{\epsilon_b}{10^{-2}}\right)
    \left(\frac{\xi_\chi}{0.03}\right)
    \left(\frac{500~{\rm GeV}}{m_\chi}\right)^2
    \left(\frac{3}{C_D\alpha_D}\right)^2 .
    \label{eq:solar-bottom-channel-estimate}
\end{equation}
For the MSSM field content alone, the gluonic contribution is
further suppressed because $C_3^{\rm eff}$ is generated by
mass-suppressed quark loops and is negligible for the
solar-neutrino constraint.  Additional heavy colored matter can
instead generate a sizable local QCD anomaly contribution and
enhance the gluonic channel.  Even for $C_3^{\rm eff}$ of order
unity, however, its contribution to the high-energy solar-neutrino
signal remains strongly suppressed by the small neutrino yield,
$\epsilon_{gg}\sim10^{-3}$.

For the MSSM reference coefficients, taking
$\epsilon_{WW}\simeq0.9$, $\epsilon_{ZZ}\simeq1.1$, and
$\epsilon_\tau\simeq1.1$, together with the order-one estimates
$\epsilon_t\simeq\epsilon_{Z\gamma}\simeq1$, the total
solar-neutrino signal for the reference parameters is approximately
one half of the current IceCube limit.  
The $t\bar t$ and $WW$ channels give comparable leading visible
contributions, while the $ZZ$ contribution is strongly suppressed
by destructive interference.  
The residual standard-Higgsino contribution is
subdominant but non-negligible.  
The reference point therefore lies
below the current solar-neutrino limit, although a more precise
comparison would require a dedicated treatment of the
channel-dependent detector response.

The destructive interference does not, however, generically
eliminate the solar-neutrino signal.  In the MSSM limit, the
cancellation points of the $WW$ and $ZZ$ channels do not coincide.
A cancellation in one electroweak channel therefore leaves a
nonvanishing contribution from the other.  In addition, the
residual standard-Higgsino contribution and the fermionic
hard-neutrino channels remain.  Increasing $\Delta_\phi$, for
example, strengthens the destructive interference while weakening
the velocity-dependent environmental enhancement, thereby
increasing the relative importance of the residual standard-Higgsino
contribution.  The solar-neutrino signal is therefore determined by
the interplay among electroweak interference, the residual standard
contribution, and the remaining visible channels, rather than by a
cancellation in any single channel.

It is also useful to isolate the constraint from the residual
standard-Higgsino contribution.  Requiring
$S_{\odot,\rm std}<s_\odot^{\rm lim}(m_\chi,\Delta m_0)$ gives
\begin{equation}
    \xi_\chi
    \lesssim
    0.1
    \left(
        \frac{s_\odot^{\rm lim}}{8\times10^{-5}}
    \right)^{1/2}
    \left(
        \frac{m_\chi}{500~{\rm GeV}}
    \right)^{3/2}
    \left(
        \frac{1.5~{\rm GeV}}{T_{\rm ann}}
    \right)^{1/2}
    \left(
        \frac{
            1+3.3\,T_{\rm ann}/\Delta_\phi
        }{
            3.5
        }
    \right)^{1/2}.
    \label{eq:solar-higgsino-fraction-bound}
\end{equation}  
This relation isolates the effect of the environmental enhancement
and provides a useful reference for the allowed Higgsino fraction.
When the axion-induced channel contributions are nonnegative, it is
a necessary condition for satisfying the full IceCube constraint.
With sufficiently strong destructive interference, however, an
individual channel contribution can become negative and partially
cancel the corresponding standard-Higgsino contribution.  In this
case, the constraint on $\xi_\chi$ must instead be evaluated from
the complete neutrino-weighted signal and can be weaker than the
estimate above.

For the MSSM reference coefficients, the solar-neutrino signal
receives comparable contributions from several channels rather than
being dominated by a single one.  Increasing $\tan\beta$ suppresses
the $t\bar t$ contribution, while the electroweak gauge-boson
channels remain sensitive to the near-threshold Higgsino loop and
its interference with the standard Higgsino amplitudes.  Additional
heavy electroweak matter can modify both effects through local
contributions to $C_{1,2}^{\rm eff}(s)$, changing the magnitude and
potentially the sign of the interference.  The IceCube constraint
should therefore be evaluated using the full set of
neutrino-weighted physical annihilation rates.

Values of $C_D\alpha_D$ of order a few suppress the visible
axion-mediated channels relative to the dark-photon channel while
remaining compatible with a perturbative hidden-sector
realization.  Much larger values can place increasingly restrictive
requirements on the ultraviolet completion.  We therefore focus on
$C_D\alpha_D$ of order a few in the phenomenological analysis.

\subsection{Galactic gamma-ray signatures and constraints}
\label{subsec:gamma-ray-constraints}

The solar-neutrino analysis shows that a Higgsino fraction at the
few-percent level can remain compatible with the IceCube constraint
even in the efficient-capture regime.  Such a subdominant Higgsino
population can nevertheless produce an appreciable Galactic gamma-ray
signal because the low-velocity annihilation dynamics have
qualitatively different consequences in the Sun and the halo.  In
the Sun, capture--annihilation equilibrium fixes the total
annihilation rate, so that the enhanced axion-mediated contribution
primarily redistributes annihilations among the available final
states.  In the Galactic halo, by contrast, the signal depends
directly on the low-velocity annihilation cross section.
Environment-dependent annihilation can therefore suppress the
observable solar-neutrino signal while retaining an appreciable
Galactic gamma-ray signal.

For a visible final state $X$, the effective halo annihilation rate
relevant for indirect searches is
\begin{equation}
    \langle\sigma v\rangle_X^{\rm eff}
    =
    \xi_\chi^2
    \left[
        \langle\sigma v\rangle_{{\rm EW},X}^{\rm halo}
        +
        (1+\kappa_X)
        \langle\sigma v\rangle_{\phi,X}^{\rm halo}
    \right],
    \label{eq:gamma-effective-rate-phen}
\end{equation}
where $\kappa_X$ is the interference parameter defined in
Eq.~\eqref{eq:kappa-interference}.  For $R_{\rm halo}\gg1$, the
total low-velocity annihilation rate is dominated by the
axion-mediated contribution, so that
\begin{equation}
    \xi_\chi^2
    \langle\sigma v\rangle_{\phi,X}^{\rm halo}
    \simeq
    \xi_\chi^2 R_{\rm halo}
    \langle\sigma v\rangle_{\rm EW,\chi\chi}^{\rm halo}
    {\rm Br}(\phi\to X).
    \label{eq:gamma-rate-Rhalo}
\end{equation}
For the $WW$ and $ZZ$ channels, this contribution must be combined
with the standard Higgsino contribution and its interference as in
Eq.~\eqref{eq:gamma-effective-rate-phen}.  For the other
visible channels, the interference is subleading, so that the
axion-mediated contribution gives the leading effective halo rate.

The environmental enhancement $R_{\rm halo}$ compensates the usual
quadratic density suppression of the axion-mediated halo signal.
In the nearly degenerate Higgsino limit, the environmental relation
derived in Sec.~\ref{sec:environment} gives
\begin{equation}
    R_{\rm halo}
    \simeq
    K_0 R_{\rm cos}
    \left(
        1+3.3\frac{T_{\rm ann}}{\Delta_\phi}
    \right),
    \label{eq:halo-enhancement-phen}
\end{equation}
for $R_{\rm cos}\gg1$.  Here $K_0\simeq12.4$ accounts for the
change from the full Higgsino multiplet during cosmological
reannihilation to the lightest neutral state in the present halo,
while the factor in parentheses describes the additional
velocity-dependent enhancement from the subthreshold mediator.
For given $T_{\rm ann}$, $m_\chi$, and $\xi_\chi$, the cosmological
enhancement $R_{\rm cos}$ is fixed by the relic-abundance condition
in Eq.~\eqref{eq:Rcos-Tann-xi}.

Combining these relations with
$\langle\sigma v\rangle_{\rm EW,\chi\chi}^{\rm halo}
\propto m_\chi^{-2}$ gives the characteristic scaling
\begin{equation}
    \xi_\chi^2
    \langle\sigma v\rangle_{\phi,X}^{\rm halo}
    \propto
    \frac{\xi_\chi m_\chi}{T_{\rm ann}}\,
    \left(
        1+3.3\frac{T_{\rm ann}}{\Delta_\phi}
    \right)
    {\rm Br}(\phi\to X),
    \label{eq:halo-rate-relic-scaling}
\end{equation}
up to the mild mass dependence of the electroweak reference rate
and the channel branching fractions.  The relic-abundance condition
thus compensates one power of the usual $\xi_\chi^2$ density
suppression, so that the axion-mediated Galactic signal scales
approximately linearly with the Higgsino fraction.  Gamma-ray
searches can consequently remain sensitive even when the Higgsino
constitutes only a few percent of the dark matter.  
This scaling should not be extrapolated to arbitrarily small
$\xi_\chi$.  As shown in Eq.~\eqref{eq:fphi-relic-estimate}, a
smaller Higgsino fraction requires a smaller $f_\phi$, thereby
increasing the mediator width.  Finite-width corrections remain
subleading in the parameter region considered here, so that the
narrow-width approximation is adequate.  For smaller $f_\phi$,
however, finite-width effects become increasingly important and the
full thermal average should be used.

In the dark-width-dominated regime, the leading continuum rates can
be estimated as follows.  For the $WW$ and $ZZ$ channels, we include
the standard Higgsino contribution together with the axion-mediated
contribution and its interference, while the corresponding standard
contributions and interference effects are subleading for the
remaining channels.
We then obtain
\begin{align}
\frac{\langle\sigma v\rangle_{WW}^{\rm eff}}
     {\langle\sigma v\rangle_{\rm ref}}
&\simeq
0.19
\left(\frac{\xi_\chi}{0.03}\right)^2
\left(\frac{500~{\rm GeV}}{m_\chi}\right)^2
+
1.95\,{\cal E}_0\,(1+\kappa_{WW})
\left(\frac{C_2^{\rm eff}}{2.47}\right)^2
\left(\frac{3}{C_D\alpha_D}\right)^2 ,
\nonumber\\[1mm]
\frac{\langle\sigma v\rangle_{ZZ}^{\rm eff}}
     {\langle\sigma v\rangle_{\rm ref}}
&\simeq
0.16
\left(\frac{\xi_\chi}{0.03}\right)^2
\left(\frac{500~{\rm GeV}}{m_\chi}\right)^2
\nonumber\\
&\quad+
0.67\,{\cal E}_0\,(1+\kappa_{ZZ})
\left(
    \frac{C_2^{\rm eff}+0.15\,C_1^{\rm eff}}{2.69}
\right)^2
\left(\frac{3}{C_D\alpha_D}\right)^2 ,
\nonumber\\[1mm]
\frac{\langle\sigma v\rangle_{gg}^{\rm eff}}
     {\langle\sigma v\rangle_{\rm ref}}
&\simeq
9.8\times10^{-4}\,
{\cal E}_0
\left(\frac{C_3^{\rm eff}}{10^{-2}}\right)^2
\left(\frac{3}{C_D\alpha_D}\right)^2 ,
\nonumber\\
\frac{\langle\sigma v\rangle_{t\bar t}^{\rm eff}}
     {\langle\sigma v\rangle_{\rm ref}}
&\simeq
0.68\,{\cal E}_0
\left(\frac{500~{\rm GeV}}{m_\chi}\right)^2
\left(\frac{8}{\tan\beta}\right)^4
\left(\frac{3}{C_D\alpha_D}\right)^2 ,
\nonumber\\
\frac{\langle\sigma v\rangle_{b\bar b}^{\rm eff}}
     {\langle\sigma v\rangle_{\rm ref}}
&\simeq
0.84\,{\cal E}_0
\left(\frac{500~{\rm GeV}}{m_\chi}\right)^2
\left(\frac{3}{C_D\alpha_D}\right)^2 ,
\nonumber\\
\frac{\langle\sigma v\rangle_{\tau\tau}^{\rm eff}}
     {\langle\sigma v\rangle_{\rm ref}}
&\simeq
0.16\,{\cal E}_0
\left(\frac{500~{\rm GeV}}{m_\chi}\right)^2
\left(\frac{3}{C_D\alpha_D}\right)^2 .
\label{eq:halo-continuum-rates}
\end{align}
Here $\langle\sigma v\rangle_{\rm ref}
\equiv10^{-28}~{\rm cm^3\,s^{-1}}$ sets the normalization, and
\begin{equation}
    {\cal E}_0
    \equiv
    \left(\frac{\xi_\chi}{0.03}\right)
    \left(\frac{m_\chi}{500~{\rm GeV}}\right)
    \left(\frac{1.5~{\rm GeV}}{T_{\rm ann}}\right)
    \left(
        \frac{
            1+3.3\,T_{\rm ann}/\Delta_\phi
        }{
            3.5
        }
    \right).
    \label{eq:halo-common-normalization}
\end{equation}
The factor ${\cal E}_0$ collects the common dependence of the
axion-mediated halo rates that follows from the relic-abundance
condition and the environmental enhancement, and is of order unity
over the parameter region of interest.  For the fermionic channels,
we have used $c_t=\cos^2\beta$ and
$c_b=c_\tau=\sin^2\beta$, with the normalization
$q_\phi(H_u)+q_\phi(H_d)=1$.  For moderate to large $\tan\beta$,
$c_b$ and $c_\tau$ are close to unity and have been absorbed into
the numerical coefficients, while the top-quark coupling gives the
explicit $\tan\beta$ dependence.
The electroweak coefficients are evaluated at the annihilation
virtuality $q^2=s$, with the MSSM reference values taken in the
zero-velocity limit, $s\simeq4m_\chi^2$.
In the MSSM,
$C_3^{\rm eff}\sim10^{-2}$ represents the typical size of the
mass-suppressed quark-loop contribution.
Electroweak Sommerfeld corrections are not included in these estimates~\cite{Chun:2015mka,Fukuda:2024dvi}.  They can modify the low-velocity annihilation rates and
the residual $ZZ$ contribution near the cancellation point, but do not
qualitatively affect the overall phenomenology discussed below.

For the MSSM field content and $\tan\beta=8$, the $WW$,
$b\bar b$, and $t\bar t$ channels give comparable leading
continuum contributions. 
Destructive interference substantially reduces the $WW$ rate
relative to the incoherent sum of the standard and axion-mediated
contributions and suppresses the axion-induced contribution to the
$ZZ$ channel, leaving the latter close to its standard Higgsino
value.
The $t\bar t$ contribution decreases rapidly with
increasing $\tan\beta$, whereas the $WW$ and $ZZ$ channels do not
share this suppression.  The gluonic continuum remains strongly
suppressed in the MSSM because of the small
$C_3^{\rm eff}$.

The standard thermal Higgsino provides a useful reference for the
current sensitivity of Galactic gamma-ray searches.  For
$m_\chi\simeq1.1~{\rm TeV}$ and $\xi_\chi=1$, recent H.E.S.S. line
searches have begun to probe the predicted annihilation signal for
cuspy Galactic dark-matter profiles~\cite{HESS:2026ila}.  In the
present scenario, the Higgsino constitutes only a subdominant
component of dark matter, while the environmental enhancement
compensates part of the corresponding density suppression.
Galactic gamma-ray searches can therefore remain relevant even for
a Higgsino fraction at the few-percent level.

We compare the predicted halo rates with the observed 95\% C.L.
upper limits from the H.E.S.S. Inner Galaxy Survey (IGS)~\cite{HESS:2022IGS,HESS:2026ila}, adopting the
baseline Einasto profile and assuming that the Higgsino component
traces the Galactic dark-matter distribution.  Among the standard
continuum final states, H.E.S.S. is most sensitive to
$\tau^+\tau^-$ over the mass range of interest because of its
relatively hard gamma-ray spectrum.  The $WW$ channel provides the
next strongest constraint, while the limits on $ZZ$, $b\bar b$,
and $t\bar t$ are weaker.  For the MSSM reference coefficients,
the continuum rates in Eq.~\eqref{eq:halo-continuum-rates},
including the $WW$ and $ZZ$ interference effects, remain well below
the current H.E.S.S. limits.

A qualitatively different continuum signal can arise in the
presence of additional heavy colored matter.  If such states
generate an order-one local QCD anomaly contribution, the
mass-suppressed MSSM contribution to $C_3^{\rm eff}$ becomes
subdominant.  For $C_3^{\rm eff}$ of order unity, the gluonic
channel becomes the dominant visible continuum mode, with a rate
about an order of magnitude larger than the leading MSSM continuum
channels for the reference parameters.

Since H.E.S.S. does not provide a dedicated limit on the $gg$
final state, we estimate the corresponding constraint by recasting
the published IGS continuum limits.  We generate prompt gamma-ray
spectra for $gg$ and the five reference channels $WW$, $ZZ$,
$b\bar b$, $t\bar t$, and $\tau^+\tau^-$ using
CosmiXs~\cite{Arina:2023CosmiXs} within a common shower and
hadronization framework.  We then reconstruct an energy-binned
non-negative effective response from the published H.E.S.S. limits,
using smoothness regularization, and validate the procedure by
successively removing each reference channel from the
reconstruction.  For $m_\chi=0.5$--$2~{\rm TeV}$, the resulting
limit on the $gg$ channel ranges from approximately
$5.8\times10^{-25}~{\rm cm^3\,s^{-1}}$ at
$m_\chi=0.5~{\rm TeV}$ to
$1.1\times10^{-25}~{\rm cm^3\,s^{-1}}$ at
$m_\chi=2~{\rm TeV}$.  We estimate the uncertainty associated with
the spectral recast to be about $25\%$ over this mass range.  At
$m_\chi\simeq0.3~{\rm TeV}$, only the high-energy tail of the
gluonic spectrum lies within the H.E.S.S. energy range, making the
recast substantially less robust.  We therefore restrict the
quantitative comparison to $m_\chi\gtrsim0.5~{\rm TeV}$.  At lower
masses, dwarf-spheroidal gamma-ray searches provide complementary
constraints with greater sensitivity to the lower-energy part of
the continuum spectrum.

For $C_3^{\rm eff}$ of order unity, the predicted gluonic rate
remains below the current H.E.S.S. sensitivity over the parameter
region of interest, although the separation decreases for larger
${\cal E}_0$ or smaller $C_D\alpha_D$.  A larger local QCD anomaly
contribution can further enhance the signal, since the gluonic rate
scales as
${\cal E}_0(C_3^{\rm eff}/C_D\alpha_D)^2$.  The gluonic channel is
particularly complementary to the solar-neutrino signal because
its high-energy neutrino yield is strongly suppressed, whereas
hadronization and subsequent decays produce an appreciable
continuum gamma-ray spectrum.  It therefore provides a
complementary probe of additional heavy colored matter.

The near-threshold Higgsino loop also generates the
$\gamma\gamma$ and $Z\gamma$ line signals.  To compare with the
H.E.S.S. line search, we define the effective line rate as
\begin{equation}
    \langle\sigma v\rangle_{\rm line}^{\rm eff}
    \equiv
    \langle\sigma v\rangle_{\gamma\gamma}^{\rm eff}
    +
    \frac{1}{2}
    \langle\sigma v\rangle_{Z\gamma}^{\rm eff},
    \label{eq:line-effective}
\end{equation}
where the factor $1/2$ accounts for the single line photon in the
$Z\gamma$ final state relative to the two photons in the
$\gamma\gamma$ final state.  In the dark-width-dominated regime,
the effective line rate can be estimated as
\begin{equation}
    \frac{
        \langle\sigma v\rangle_{\rm line}^{\rm eff}
    }{
        \langle\sigma v\rangle_{\rm ref}
    }
    \simeq
    0.3\,{\cal E}_0
    \left[
        \frac{
            (C_2^{\rm eff})^2
            +0.83\,(C_1^{\rm eff})^2
        }{
            7.9
        }
    \right]
    \left(\frac{3}{C_D\alpha_D}\right)^2 .
    \label{eq:halo-line-rate}
\end{equation}
For simplicity, the $Z\gamma$ phase-space factor
$(1-m_Z^2/m_\phi^2)^3$ has been neglected in this estimate.  The
cross term proportional to $C_1^{\rm eff}C_2^{\rm eff}$ nearly
cancels between the $\gamma\gamma$ and $Z\gamma$ contributions to
the effective line rate, leaving only a weak dependence on the
relative sign of the two electroweak coefficients.  For the MSSM
reference coefficients evaluated at $q^2=s\simeq4m_\chi^2$, the
factor in parentheses is approximately unity.  The line signal is
therefore directly related to the same matched electroweak
coefficients that determine the axion-mediated $WW$ and $ZZ$
amplitudes.

Although the predicted line rate is smaller than the leading
continuum rates, H.E.S.S. has substantially greater sensitivity to
a spectrally localized feature.  For the baseline Einasto profile,
the observed line limit ranges from approximately
$1.1\,\langle\sigma v\rangle_{\rm ref}$ at
$m_\chi=0.5~{\rm TeV}$ to
$5.8\,\langle\sigma v\rangle_{\rm ref}$ at
$m_\chi=2~{\rm TeV}$~\cite{HESS:2026ila}.  At
$m_\chi=500~{\rm GeV}$, the MSSM reference prediction is
approximately $0.30\,\langle\sigma v\rangle_{\rm ref}$ for
$C_D\alpha_D=3$, only about a factor of four below the current
H.E.S.S. limit.  Thus, even when the Higgsino constitutes only a
few percent of the dark matter, the predicted line signal can lie
close to the current experimental sensitivity.  This is a direct
consequence of the low-velocity enhancement, which compensates part
of the quadratic density suppression of the subdominant Higgsino
component.  The line signal increases further with ${\cal E}_0$
and with decreasing $C_D\alpha_D$, and can approach the current
H.E.S.S. sensitivity within the parameter region of interest.

The gamma-ray line signal is closely related to the solar-neutrino
constraint through the same matched electroweak coefficients
$C_{1,2}^{\rm eff}$.  These coefficients determine the
axion-mediated electroweak amplitudes and their interference with
the standard Higgsino amplitudes in the solar signal, while also
setting the $\gamma\gamma$ and $Z\gamma$ line rates.  The parameter
region relevant for the current IceCube sensitivity can therefore
also yield a Galactic line signal close to the present H.E.S.S.
sensitivity.

The two line channels probe different combinations of the matched
electroweak coefficients.  The $\gamma\gamma$ amplitude is
proportional to $C_2^{\rm eff}+1.7\,C_1^{\rm eff}$, whereas the
$Z\gamma$ amplitude is proportional to
$C_2^{\rm eff}-0.5\,C_1^{\rm eff}$.  Additional heavy electroweak
matter can therefore modify both the overall line strength and the
relative importance of the two channels.  For the MSSM field
content, by contrast, their relative strengths follow from the
characteristic relation between $C_1^{\rm eff}$ and
$C_2^{\rm eff}$ generated by the near-threshold Higgsino loop.

Taken together, the continuum and line searches probe complementary
aspects of the Galactic signal.  For the MSSM field content,
destructive interference reduces the electroweak continuum rates,
which remain well below the current H.E.S.S. limits, while the
$\gamma\gamma$ and $Z\gamma$ lines provide the more sensitive
Galactic probe.  Additional heavy colored matter can qualitatively
change the continuum phenomenology by generating an order-one local
QCD anomaly contribution and making the gluonic channel dominant.
The environmental enhancement can therefore leave observable
Galactic signatures even when the Higgsino constitutes only a few
percent of the dark matter, with gamma-ray line searches providing
sensitivity close to the parameter region considered here.


\subsection{Terrestrial inelastic scattering}
\label{subsec:terrestrial-inelastic}

The suppression of the solar-neutrino signal has an important
implication for terrestrial searches.  The off-diagonal $Z$
interaction responsible for inelastic solar capture also induces
up-scattering of the lightest neutral Higgsino on nuclei,
\begin{equation}
    \chi N\to\chi_2^0 N.
    \label{eq:terrestrial-inelastic-process}
\end{equation}
Unlike the solar-neutrino signal, however, the terrestrial recoil
rate is insensitive to the subsequent annihilation dynamics.
Assuming that the local Higgsino fraction follows its cosmological
abundance fraction, the differential event rate can be written
schematically as
\begin{equation}
    \frac{dR_{\rm DD}}{dE_R}
    =
    \xi_\chi\frac{\rho_{\rm DM}}{m_\chi}
    \frac{d\langle\sigma_{\rm inel}v\rangle}{dE_R},
    \label{eq:direct-detection-rate-scaling}
\end{equation}
and therefore scales linearly with $\xi_\chi$ for fixed $m_\chi$,
$\Delta m_0$, and the Galactic velocity distribution.
The neutral-Higgsino splitting controls the kinematic accessibility
of the inelastic transition.  For scattering on a nucleus of mass
$m_N$, the minimum incoming velocity required to produce a recoil
energy $E_R$ is~\cite{Tucker-Smith:2001myb}
\begin{equation}
    v_{\rm min}(E_R)
    =
    \frac{1}{\sqrt{2m_NE_R}}
    \left(
        \frac{m_NE_R}{\mu_{\chi N}}
        +\Delta m_0
    \right),
    \label{eq:inelastic-vmin}
\end{equation}
where $\mu_{\chi N}$ is the Higgsino--nucleus reduced mass.  The
minimum of $v_{\rm min}$ occurs at
\begin{equation}
    E_R^*
    =
    \frac{\mu_{\chi N}}{m_N}\Delta m_0,
    \qquad
    v_{\rm thr}
    =
    \sqrt{\frac{2\Delta m_0}{\mu_{\chi N}}}.
    \label{eq:inelastic-threshold}
\end{equation}
A neutral-state splitting of a few hundred keV therefore selects
both large recoil energies and the high-velocity tail of the
Galactic Higgsino distribution.  For a several-hundred-GeV Higgsino
scattering on xenon, $E_R^*$ lies naturally in the few-hundred-keV
range, making high-recoil searches particularly sensitive to the
quasi-Dirac regime considered here.

The neutral-state splitting $\Delta m_0$ also controls inelastic
capture in the Sun, but the observable consequences are qualitatively
different.  Once capture--annihilation equilibrium is established,
the solar-neutrino signal scales as
\begin{equation}
    S_\odot
    =
    \xi_\chi F_{\nu,\odot},
    \label{eq:solar-direct-comparison}
\end{equation}
whereas the terrestrial recoil rate retains the direct scaling
$R_{\rm DD}\propto\xi_\chi$.  Environment-dependent annihilation can
make $F_{\nu,\odot}\ll1$ after capture without modifying the
$Z$-mediated scattering process.  Suppressing the solar signal
through the annihilation dynamics therefore preserves a much larger
terrestrial rate than suppressing it by reducing the Higgsino
abundance alone.

This distinction is particularly relevant in the few-hundred-keV
splitting regime, where inelastic solar capture can remain efficient
while terrestrial scattering is kinematically accessible only to the
high-velocity part of the Galactic distribution.  
A Higgsino fraction at the few-percent level can remain compatible
with the solar-neutrino constraint because the environmental
enhancement suppresses the residual standard-Higgsino contribution,
while the mediator decay pattern and destructive interference reduce
the hard-neutrino signal from the visible channels.  The terrestrial
recoil rate then retains its linear dependence on $\xi_\chi$ and can
remain appreciable.

The recent LZ high-recoil event provides a direct application of
this mechanism~\cite{LZ:2026axp}.  Inelastic Higgsino scattering
with a neutral-state splitting of a few hundred keV populates the
high-recoil region~\cite{Freese:2026sga}, while the event rate can
remain appreciable for a subdominant Higgsino component.
The present scenario therefore provides a framework for interpreting
the LZ event in terms of quasi-Dirac Higgsino scattering while
alleviating the otherwise severe solar-neutrino constraint through
environment-dependent annihilation.  A dedicated fit to the LZ recoil
spectrum and detector response would be required to determine the
preferred Higgsino mass, splitting, and abundance fraction.


\subsection{Combined phenomenology}
\label{subsec:combined-phenomenology}

The effect of environment-dependent annihilation is most transparent
when compared with a Higgsino scenario without axion-mediated
annihilation, at fixed $m_\chi$, $\Delta m_0$, and $\xi_\chi$.
For the solar-neutrino signal, the axion-mediated
Galactic-halo annihilation signal, and terrestrial inelastic
scattering, the parametric dependence changes schematically as
\begin{equation}
\begin{aligned}
    S_\odot &: \qquad
    \xi_\chi
    \;\longrightarrow\;
    \xi_\chi F_{\nu,\odot},
    \\ 
    \langle\sigma v\rangle_{\phi,X}^{\rm eff}
    &: \qquad
    \xi_\chi^2
    \;\longrightarrow\;
    \xi_\chi^2 R_{\rm halo},
    \\
    R_{\rm DD} &: \qquad
    \xi_\chi
    \;\longrightarrow\;
    \xi_\chi .
\end{aligned}
\label{eq:combined-scaling}
\end{equation}
In the Sun, the neutrino signal acquires the additional suppression
$F_{\nu,\odot}$ from the modified annihilation pattern,
including the redistribution into neutrino-poor final states and
destructive interference in the visible electroweak channels.  
In the Galactic halo, the environmental enhancement compensates one power of the
usual $\xi_\chi^2$ density suppression of the
axion-mediated contribution once the relic-abundance condition is
imposed.  For the $WW$ and $ZZ$ channels, the physical
rates additionally contain the standard Higgsino contribution and
its interference with the axion-mediated amplitude, as discussed
above. The terrestrial inelastic-scattering rate, by contrast, is
unaffected by the annihilation dynamics and remains directly
proportional to the local Higgsino abundance.

These different dependences are the central phenomenological feature
of our scenario.  If the solar-neutrino constraint were evaded solely
by reducing the Higgsino abundance, the terrestrial scattering rate
would decrease linearly with $\xi_\chi$, while the standard Galactic
annihilation signal would decrease quadratically.  Environment-dependent
annihilation instead suppresses the observable solar-neutrino signal
after capture without modifying the $Z$-mediated terrestrial scattering
process, while the enhanced axion-mediated contribution
can retain an appreciable Galactic gamma-ray signal.  A Higgsino
fraction at the few-percent level can therefore remain compatible with
the solar-neutrino constraint while retaining appreciable Galactic and
terrestrial signals.

The parameters controlling these observables have distinct physical
roles.  The neutral-state splitting $\Delta m_0$ determines the
kinematics of both solar capture and terrestrial inelastic
scattering.  The reannihilation temperature $T_{\rm ann}$ and the
ratio $\Delta_\phi/T_{\rm ann}$ control the relation between the
cosmological and low-velocity annihilation rates.  The combination
$C_D\alpha_D$ sets the invisible mediator width and hence the
competition between hidden and visible final states, while the
matched coefficients $C_i^{\rm eff}$ and $\tan\beta$ determine
the visible annihilation pattern, including the
continuum and gamma-ray line channels as well as interference in
the electroweak gauge-boson channels.  The different searches
therefore probe complementary combinations of the underlying
parameters rather than a single annihilation strength.

\begin{figure}[t]
  \centering
  \includegraphics[width=0.72\linewidth]{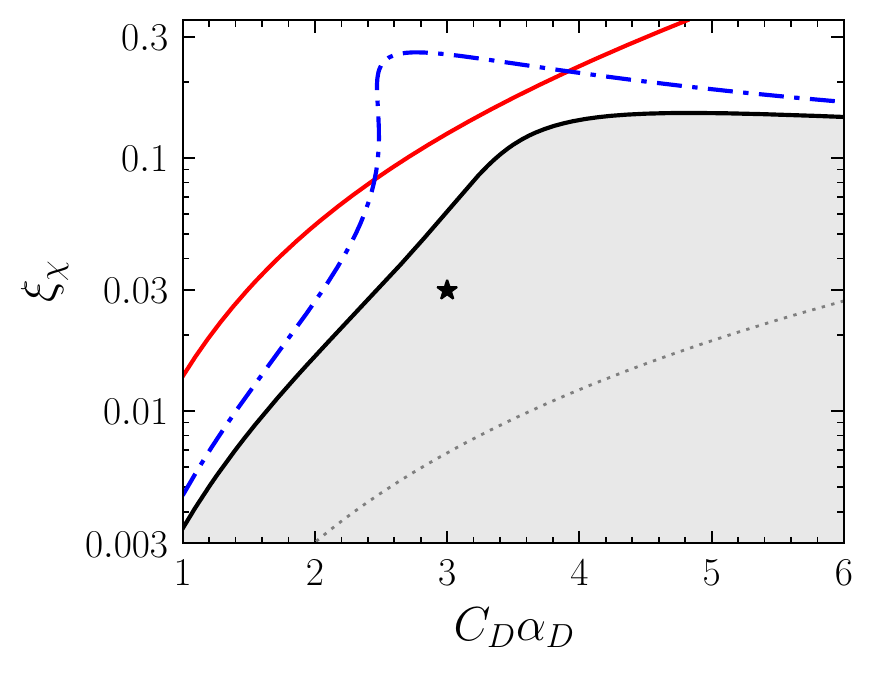}
  \caption{
    Solar-neutrino and gamma-ray line constraints in the
    $(C_D\alpha_D,\xi_\chi)$ plane for
    $m_\chi=500\,{\rm GeV}$,
    $\Delta m_0=320\,{\rm keV}$,
    $T_{\rm ann}=1.5\,{\rm GeV}$, and
    $\Delta_\phi=2\,{\rm GeV}$.
    The black solid curve shows the solar-neutrino boundary
    $S_\odot=s_\odot^{\rm lim}$ for the MSSM reference coefficients
    with $\tan\beta=8$, using
    $s_\odot^{\rm lim}=8\times10^{-5}$.
    The blue dot-dashed curve shows the solar-neutrino boundary
    in the fermion-suppressed limit with the Higgsino-induced electroweak
    coefficients unchanged.
    The red solid curve shows the boundary
    $\langle\sigma v\rangle_{\rm line}^{\rm eff}
    =\langle\sigma v\rangle_{\rm line}^{\rm lim}$,
    using the H.E.S.S. baseline-Einasto limit
    $\langle\sigma v\rangle_{\rm line}^{\rm lim}
    \simeq1.1\times10^{-28}\,{\rm cm^3\,s^{-1}}$.
    The light-gray region satisfies both limits for the MSSM reference.
    The gray dotted curve denotes $\Gamma_\phi/\Delta_\phi=1$.
    The line and width contours are evaluated for the MSSM reference;
    the line contour also provides an approximate reference
    for the fermion-suppressed limit.
    The star marks
    $(C_D\alpha_D,\xi_\chi)=(3,0.03)$.
  }
  \label{fig:higgsino-plane}
\end{figure}

Figure~\ref{fig:higgsino-plane} illustrates these relations in the
$(C_D\alpha_D,\xi_\chi)$ plane.
At each point, the relic-abundance condition in Eq.~(\ref{eq:fphi-relic-condition}) fixes
$f_\phi$, with the thermal integral in Eq.~(\ref{eq:thermal-average-threshold}) evaluated
numerically including the finite mediator width in the fully
degenerate coannihilation limit.
For the MSSM reference coefficients, the solar-neutrino constraint
is stronger than the H.E.S.S. line constraint over the range shown,
while the reference point with $\xi_\chi=0.03$ satisfies both limits.
Suppressing the direct SM-fermion couplings, as in the extended
Higgs-sector realization discussed in Sec.~\ref{subsec:extended-higgs}, relaxes the
solar-neutrino constraint while retaining the Higgsino-induced
electroweak couplings.  Destructive interference in the $WW$ and
$ZZ$ channels further increases the allowed $\xi_\chi$ by suppressing
the remaining hard-neutrino signal.  The gamma-ray line signal remains
nearly unchanged, allowing line searches to test regions that remain
allowed by solar neutrinos.

The few-hundred-keV splitting regime provides a particularly
interesting realization of this combined phenomenology.  The range
of splittings relevant for interpreting the recent LZ high-recoil
event in terms of quasi-Dirac Higgsino scattering also permits
efficient solar capture and hence leads to a stringent
solar-neutrino constraint.  Environment-dependent annihilation can
alleviate this tension while retaining a subdominant Higgsino
component and an appreciable terrestrial recoil rate.  The
associated Galactic gamma-ray signal provides a complementary
probe.  For the MSSM field content, destructive
interference suppresses the electroweak continuum, while the
$\gamma\gamma$ and $Z\gamma$ line signals can approach the current
sensitivity even when the Higgsino constitutes only a few percent
of the dark matter.  Additional heavy colored matter can
qualitatively modify the continuum signal by generating
an order-one local QCD anomaly contribution and making the gluonic
channel dominant.

The Galactic-centre comparison is subject to the usual uncertainty
in the inner Milky-Way dark-matter profile.  Dwarf spheroidal
galaxies provide complementary gamma-ray constraints with different
astrophysical systematics and are particularly useful toward the
lower end of the Higgsino-mass range, where the H.E.S.S.
Galactic-centre sensitivity deteriorates~\cite{FermiLAT:2026Dwarfs}.
In the parameter region considered here, Higgsino annihilation in
both environments probes the effectively zero-velocity limit, so
that the same particle-physics annihilation rates apply.

Taken together, the solar, Galactic, and terrestrial signals provide
complementary probes of the same underlying Higgsino sector.
Solar-neutrino searches probe the neutrino-weighted annihilation
pattern after capture, Galactic gamma-ray observations probe the
low-velocity visible annihilation rates, and terrestrial experiments
probe the local Higgsino population through inelastic scattering.
Environment-dependent annihilation allows these observables to have
parametrically different strengths while retaining a common
particle-physics origin.  A suppressed solar-neutrino signal
accompanied by appreciable Galactic gamma-ray and terrestrial
inelastic-scattering signals would therefore constitute a
characteristic phenomenological signature of the scenario.


\section{Supersymmetric realization}
\label{sec:susy-realization}

Supersymmetry at scales above the electroweak scale remains an
interesting possibility in view of the stringent collider limits on
conventional weak-scale spectra.  A superparticle spectrum in the
multi-ten-TeV range can evade these limits while allowing lighter
electroweak states to retain observable dark-matter phenomenology.
In this regime, gauge coupling unification and the observed
Higgs-boson mass provide important constraints on the supersymmetric
spectrum.

The scenario considered here can be embedded in such a
supersymmetric framework.  The Higgsino remains near the weak scale
while the supersymmetric mass scale is much higher, with its neutral
components forming a quasi-Dirac pair with a small mass splitting
relevant for inelastic scattering.  The nonthermal cosmology involves
late Higgsino production followed by reannihilation at a temperature
of order GeV.  The axion-like mediator requires a lower
$U(1)_\phi$-breaking scale, with $f_\phi$ not far above the Higgsino
mass, together with a sizable coupling to the hidden gauge sector.
We show below that these features can arise naturally within
KKLT-type compactifications~\cite{Kachru:2003aw} with generalized mirage mediation,
supplemented by hidden strong dynamics.

\subsection{Generalized mirage mediation and the Higgsino spectrum}
\label{subsec:mirage-realization}

Generalized mirage mediation provides a hierarchy between the
gravitino mass and the soft supersymmetry-breaking masses.  In
KKLT-type compactifications, the modulus- and anomaly-mediated soft
masses are loop suppressed relative to
$m_{3/2}$~\cite{Choi:2005uz}, while a conventionally generated
Higgs bilinear parameter is generically of order $m_{3/2}$.
The regime relevant for the present scenario is therefore characterized by
\begin{equation}
    |\mu|
    \ll
    m_{\rm SUSY}
    \sim
    \frac{m_{3/2}}{8\pi^2}
    \ll
    |B|
    \sim
    m_{3/2},
    \label{eq:mirage-scale-hierarchy}
\end{equation}
where $m_{\rm SUSY}$ denotes the typical scale of the
modulus- and anomaly-mediated soft masses rather than a common
superparticle mass.  The small Higgsino mass follows from
electroweak symmetry breaking in the presence of the large Higgs
bilinear parameter.  For
$m_{H_u}^2,m_{H_d}^2\sim m_{\rm SUSY}^2$ and moderate
$\tan\beta$, the Higgs minimization condition gives parametrically
$B\mu\sim m_{\rm SUSY}^2\sin2\beta$, up to factors of order unity.
For a few-PeV gravitino, $m_{\rm SUSY}$ lies in the multi-ten-TeV
range while the Higgsino can remain near the weak scale, as required
for the phenomenology considered here~\cite{Jeong:2011sg}.

A central motivation for the KKLT realization of generalized mirage
mediation in the present scenario is cosmological.  In KKLT-type
compactifications, the lightest K\"ahler modulus is typically heavier
than the gravitino by a logarithmic factor~\cite{Choi:2005hd},
\begin{equation}
    m_T
    \sim
    m_{3/2}
    \ln\left(\frac{M_{\rm Pl}}{m_{3/2}}\right),
    \label{eq:kklt-modulus-mass}
\end{equation}
and its decay generically produces gravitino pairs with a branching
fraction at the percent level or larger~\cite{Endo:2006zj,Nakamura:2006uc,Dine:2006ii,Endo:2006tf}.
The gravitinos subsequently decay well after thermal Higgsino
freeze-out, providing a late nonthermal source of Higgsinos.  For
the gravitino abundance expected from modulus decay, the injected
Higgsino yield exceeds the critical yield for reannihilation by
several orders of magnitude, while the gravitino energy density can
remain subdominant at the time of decay.  The injected Higgsinos
therefore enter the reannihilation regime, with the annihilation
temperature set approximately by the gravitino decay temperature,
\begin{equation}
    T_{\rm ann}
    \simeq
    T_{3/2}
    =
    \left(
        \frac{90}{\pi^2 g_*(T_{3/2})}
    \right)^{1/4}
    \sqrt{\Gamma_{3/2}M_{\rm Pl}}
    \simeq
    (1.2\text{--}1.5)~{\rm GeV}
    \left(
        \frac{m_{3/2}}{4~{\rm PeV}}
    \right)^{3/2}.
    \label{eq:Tann-gravitino}
\end{equation}
The numerical coefficient depends mildly on the particle content,
since additional kinematically accessible states can modify the
total gravitino width.  A few-PeV gravitino nevertheless gives an
annihilation temperature of order GeV.  In the reannihilation regime,
the final Higgsino abundance is controlled primarily by the
annihilation rate at $T_{\rm ann}$ and is largely insensitive to
the initial gravitino abundance and the details of the preceding
modulus-decay history. See Ref.~\cite{Jeong:2012en} for a cosmological scenario with an abundant gravitino population.

The gaugino-mass pattern is comparatively predictive in generalized
mirage mediation, since it is controlled primarily by the gauge
kinetic functions and the relative contributions of modulus and
anomaly mediation.  In contrast, the scalar soft terms depend more
sensitively on the matter K\"ahler metrics and are therefore more
model dependent.  Defining
\begin{equation}
    \alpha
    \equiv
    \frac{m_{3/2}}
         {M_0\ln(M_{\rm Pl}/m_{3/2})},
    \label{eq:mirage-alpha}
\end{equation}
where $M_0$ is the modulus-mediated contribution to the gaugino
masses, the one-loop gaugino masses at a renormalization scale $Q$
satisfy~\cite{Choi:2005hd}
\begin{equation}
    \frac{M_a(Q)}{g_a^2(Q)}
    =
    \frac{M_0}{g_{\rm GUT}^2}
    \left[
        1+
        \frac{b_a g_{\rm GUT}^2}{16\pi^2}\,
        \alpha
        \ln\!\left(\frac{M_{\rm Pl}}{m_{3/2}}\right)
    \right],
    \label{eq:mirage-gaugino-pattern}
\end{equation}
where $b_a=(33/5,\,1,\,-3)$ are the MSSM one-loop gauge
beta-function coefficients.  
In generalized KKLT compactifications, the leading value of $\alpha$
is determined by moduli stabilization and the gauge kinetic functions
and can take simple rational values of order unity.
As shown below, values of $\alpha$ near the relevant cancellation
region can generate the bino--wino mass relation required for a small
neutral-Higgsino splitting.


For $|M_{1,2}|\gg|\mu|,m_Z$, the neutral-Higgsino mass splitting
can be expanded as
\begin{equation}
    \Delta m_0
    =
    \delta m_0^{(1)}
    +
    \delta m_0^{(2)}
    +\cdots ,
    \label{eq:higgsino-splitting-expansion}
\end{equation}
where the leading and next-to-leading contributions are
\begin{align}
    \delta m_0^{(1)}
    &=
    m_Z^2
    \left(
        \frac{s_W^2}{M_1}
        +
        \frac{c_W^2}{M_2}
    \right),
    \nonumber\\
    \delta m_0^{(2)}
    &=
    \mu m_Z^2\sin2\beta
    \left(
        \frac{s_W^2}{M_1^2}
        +
        \frac{c_W^2}{M_2^2}
    \right).
    \label{eq:higgsino-splitting-orders}
\end{align}
The leading contribution is sensitive to the relative sign of the
bino and wino masses and can be strongly suppressed for opposite
signs.  At leading order, the cancellation occurs for
\begin{equation}
    \frac{M_1}{M_2}
    \simeq
    -\tan^2\theta_W .
    \label{eq:bino-wino-cancellation}
\end{equation}
The next-to-leading contribution depends instead on the squared
gaugino masses and is not subject to the same cancellation.  Near
the leading-order cancellation point, it becomes
\begin{equation}
    \delta m_0^{(2)}
    \simeq
    \frac{\mu m_Z^2}{M_2^2}
    \cot^2\theta_W\,
    \sin2\beta .
    \label{eq:higgsino-splitting-nlo}
\end{equation}
For a weak-scale Higgsino and electroweak gauginos at a few tens of
TeV, the next-to-leading contribution is of order MeV.  A splitting
in the few-hundred-keV range relevant for the LZ high-recoil event
can then arise slightly away from the leading-order cancellation
point, where a small negative $\delta m_0^{(1)}$ partially cancels
the positive $\delta m_0^{(2)}$.

For a gravitino mass in the PeV range, the leading-order cancellation
corresponds to $\alpha$ close to $-3$.  In generalized KKLT
compactifications, the leading value of the mirage parameter can take
simple rational values of order unity~\cite{Choi:2006xb,Jeong:2021xse}.   
The vicinity of the cancellation may therefore be parameterized, for example, by
\begin{equation}
    \alpha
    =
    -3+\frac{1}{n},
    \label{eq:mirage-rational-alpha}
\end{equation}
with $n$ a modest positive integer. 
This parametrization identifies the vicinity of the leading-order
cancellation rather than fixing the physical splitting.  Since
$\alpha$ generically receives subleading corrections of order
$1/\ln(M_{\rm Pl}/m_{3/2})$, taking
$n\gtrsim\ln(M_{\rm Pl}/m_{3/2})$ does not provide a meaningful
refinement of the leading prediction, nor is such a refinement
required to reach the phenomenologically relevant few-hundred-keV
range.  At this level, the physical splitting depends on the
residual leading contribution together with higher-order and
threshold corrections.

Near the bino--wino cancellation region, the gaugino and gravitino
masses exhibit the approximate hierarchy
\begin{equation}
    |M_1|
    \simeq
    0.3 M_2,
    \qquad
    M_3
    \simeq
    4 M_2,
    \qquad
    m_{3/2}
    \simeq
    100 M_2,
    \label{eq:mirage-mass-hierarchy}
\end{equation}
up to corrections from renormalization-group evolution and
superparticle thresholds.  
A wino mass of a few tens of TeV therefore corresponds to a bino
near the ten-TeV scale, a gluino near the hundred-TeV scale, and a
gravitino at a few PeV.\footnote{
Ref.~\cite{Du:2026guj} instead imposes the GUT-scale boundary
condition $M_1^G/M_2^G=-3/5$ to cancel the leading bino--wino
contribution to the neutral-Higgsino splitting.  The remaining
finite-gaugino terms give $\Delta m_0\simeq350~{\rm keV}$ with
$M_2\simeq120~{\rm TeV}$ for their benchmark
$\mu=1.091~{\rm TeV}$ and $\tan\beta=10$.  In the present scenario,
the cancellation region is realized through the mirage gaugino
spectrum, allowing a few-hundred-keV splitting with a bino near
$10~{\rm TeV}$ and a wino at the multi-ten-TeV scale.
}
The mirage spectrum thus links the bino--wino mass relation required
for a small neutral-Higgsino splitting to the few-PeV gravitino mass
that sets the nonthermal reannihilation temperature at the GeV scale.

Finite wino mixing also modifies the charged--neutral Higgsino mass
splitting.  For $\mu M_2>0$, the leading tree-level contribution is
\begin{equation}
    \delta(\Delta m_\pm)_{\rm tree}
    \simeq
    -\frac{m_W^2}{M_2}\sin2\beta .
    \label{eq:charged-higgsino-tree}
\end{equation}
Together with the electroweak radiative contribution~\cite{Thomas:1998wy},
the total charged--neutral splitting is typically
$0.2$--$0.3~{\rm GeV}$ in the parameter region of interest.
The relevant hierarchy is therefore
\begin{equation}
    \Delta m_0
    \ll
    \Delta m_\pm
    \lesssim
    T_{\rm ann}
    \ll
    m_\chi .
    \label{eq:reannihilation-higgsino-hierarchy}
\end{equation}
The charged and neutral Higgsino states are consequently populated
during reannihilation, whereas only the lightest neutral state is
populated in the much colder solar and Galactic-halo environments.
The mirage spectrum therefore realizes the population hierarchy
responsible for the environmental enhancement discussed in
Sec.~\ref{sec:environment}.
The compressed charged-Higgsino spectrum is also consistent with
current collider searches, which leave the several-hundred-GeV
Higgsino region considered here viable for sub-GeV charged--neutral
mass splittings~\cite{ATLAS:2025higgsino,Cheung:2026LZ}.

The supersymmetric spectrum considered here is also compatible with
precision gauge coupling unification.  For a nondegenerate MSSM
spectrum, and neglecting GUT-scale threshold corrections, the
unification condition can be written as~\cite{Krippendorf:2013dqa,Jeong:2017ncm}
\begin{equation}
    \frac{|\mu|}{m_*}
    \left(\frac{m_H}{m_*}\right)^{1/4}
    \left(\frac{M_2}{m_*}\right)^{1/3}
    \left(\frac{M_2}{M_3}\right)^{7/3}
    \left(\frac{m_{\tilde\ell}}{m_{\tilde q}}\right)^{1/4}
    \simeq
    1 ,
    \label{eq:precision-unification-condition}
\end{equation}
where $m_*\simeq2~{\rm TeV}$ denotes the reference supersymmetric
threshold corresponding to precision unification.
The unification condition is particularly sensitive to the Higgsino
mass and the gaugino hierarchy, while its dependence on the
heavy-Higgs and sfermion masses is comparatively weak.  A light
Higgsino can therefore remain compatible with precision unification
even when the other superparticles lie well above the weak scale.
In the mirage spectrum of interest, $M_3/M_2\simeq4$ shifts the
unification-compatible superparticle scale into the multi-ten-TeV
regime, consistent with the hierarchy in
Eq.~\eqref{eq:mirage-scale-hierarchy}.  Small departures from this
relation can be accommodated by GUT-scale threshold corrections.

The gaugino hierarchy also has an important implication for proton
decay.
A relatively heavy gluino lowers the gauge-unification
scale and thereby enhances dimension-six proton decay.  Combining
precision gauge coupling unification with the proton-decay constraint
gives approximately~\cite{Jeong:2017ncm}
\begin{equation}
    m_{\rm SUSY}
    \lesssim
    40~{\rm TeV}
    \left(\frac{|\mu|}{500~{\rm GeV}}\right)
    \left(\frac{M_3/M_2}{4}\right)^{-25/8},
    \label{eq:proton-decay-susy-scale}
\end{equation}
up to the dependence on the remaining superparticle thresholds and
GUT-scale corrections. 
For a weak-scale Higgsino and $M_3/M_2\simeq4$, precision
unification and proton stability therefore point to a superparticle
spectrum in the multi-ten-TeV range.  This is consistent with the
independent estimate $m_{\rm SUSY}\sim m_{3/2}/8\pi^2$ for a
few-PeV gravitino.


The observed Higgs-boson mass provides an additional constraint on
the Higgs sector.  For superparticle masses in the multi-ten-TeV
range, the MSSM prediction for $m_h\simeq125~{\rm GeV}$ generally
favors moderate values of $\tan\beta$, with the preferred range
depending on the stop masses and mixing~\cite{Giudice:2011cg}.
We do not impose this relation as a sharp constraint, since
model-dependent threshold corrections from the heavy spectrum can
modify the Higgs quartic coupling at the supersymmetry scale.
Somewhat larger values of $\tan\beta$ can therefore remain viable
and are included in the phenomenological analysis.

\subsection{Axion sector from hidden strong dynamics}
\label{subsec:hidden-strong-dynamics}

The low $U(1)_\phi$-breaking scale required in the present scenario
can have a dynamical origin.  Since
$g_{\phi\chi\chi}\sim\mu/f_\phi$, an appreciable Higgsino coupling
requires $f_\phi$ not far above the Higgsino mass.  Such a hierarchy
can arise from hidden strong dynamics.

We introduce a singlet $S$ coupled to vectorlike hidden quarks
$Q_i$ and $Q_i^c$ charged under a confining gauge group $SU(N_H)$,
with
\begin{equation}
    W 
    =
    \sum_i y_i S Q_i Q_i^c .
    \label{eq:hidden-quark-yukawa}
\end{equation}
We assume that at least one hidden-quark flavor has a Yukawa coupling
of order unity.  If the confinement scale $\Lambda_H$ lies below
the hidden-superpartner masses, the strong dynamics below
$\Lambda_H$ is effectively nonsupersymmetric.  Hidden-quark
condensation then induces a linear term for $S$,
\begin{equation}
    \Delta V
    \sim
    \Lambda_H^3 S
    +{\rm h.c.},
    \label{eq:S-hidden-tadpole}
\end{equation}
where coefficients of order unity have been suppressed.  If the
singlet scalar mass is dominated by its soft mass $m_S$, with
supersymmetric contributions subdominant, the induced expectation
value is parametrically
\begin{equation}
    \langle S\rangle
    \sim
    \frac{\Lambda_H^3}{m_S^2}.
    \label{eq:S-vev-hidden}
\end{equation} 
A TeV-scale expectation value therefore corresponds to
\begin{equation}
    \Lambda_H
    \sim
    10~{\rm TeV}
    \left(\frac{\langle S\rangle}{1~{\rm TeV}}\right)^{1/3}
    \left(\frac{m_S}{30~{\rm TeV}}\right)^{2/3}.
    \label{eq:hidden-confinement-scale}
\end{equation}
Thus, a multi-ten-TeV singlet soft mass and a confinement scale
of order $10~{\rm TeV}$ yield $\langle S\rangle$ at the TeV scale.

The physical mediator $\phi$ is the pNGB associated with the
approximate $U(1)_\phi$ symmetry and need not coincide with the
phase of $S$.  After confinement, both $\langle S\rangle$ and the
hidden-quark condensate contribute to $U(1)_\phi$ breaking, so that
the pNGB is in general a linear combination of the singlet phase and
the corresponding hidden-pion direction.  
Denoting the hidden-pion decay constant by $f_H$, the physical decay
constant is parametrically~\cite{Luty:1997fk}
\begin{equation}
    f_\phi^2
    \sim
    \langle S\rangle^2+f_H^2
    \sim
    \left(\frac{\Lambda_H^3}{m_S^2}\right)^2
    +
    \left(\frac{\Lambda_H}{4\pi}\right)^2 ,
    \label{eq:fphi-hidden}
\end{equation}
up to factors of order unity associated with the $U(1)_\phi$ charge
assignment and the normalization of the Goldstone direction.  Both
contributions to $f_\phi$ are therefore controlled by the same
hidden confinement dynamics. 
For the scales considered above, $f_\phi$ can lie near the TeV
scale, with $f_\phi\sim2\mu$ in the region relevant for the
annihilation mechanism.

The effective interactions introduced in the preceding sections are
those of the physical pNGB.  Its couplings to the Higgsino
and to $U(1)_\phi$-charged SM fermions retain the
$1/f_\phi$ dependence, while the gauge-boson couplings are determined
by the corresponding $U(1)_\phi$ anomaly coefficients.  The
phenomenology discussed above therefore applies independently of
whether $\phi$ is predominantly the phase of $S$, predominantly
hidden-pion-like, or a sizable mixture of the two.

A nonzero pNGB mass requires explicit $U(1)_\phi$ breaking.  This
can be provided by a hidden-quark flavor with a small Yukawa coupling
$y$, giving $m_q\simeq y\langle S\rangle$ after $S$ acquires its
expectation value.  The resulting pNGB mass is parametrically
\begin{equation}
    m_\phi^2 f_\phi^2
    \sim
    m_q\Lambda_H^3 .
    \label{eq:axion-mass-hidden}
\end{equation}
For $\langle S\rangle$, $f_\phi$, and $m_\phi$ near the TeV scale
and $\Lambda_H$ of order $10~{\rm TeV}$, the required Yukawa
coupling is small.  The same hidden strong dynamics can therefore
account for both $f_\phi$ and $m_\phi$ at the TeV scale.  The
subthreshold relation $m_\phi\simeq2m_\chi$, however, remains a
separate condition on the pNGB and Higgsino masses.

Some of the hidden quarks may also carry QCD color.  Such states
generate a nondecoupling local QCD anomaly contribution,
which can be of order unity and enhance the physical gluonic
coupling considered in the gamma-ray phenomenology.
Below the hidden confinement scale,
$\Lambda_H\sim10~{\rm TeV}$, the elementary hidden quarks are
replaced by hidden hadronic states, while above this scale they
modify the QCD running and hence the precision-unification condition. 
If the colored hidden quarks are singlets under
$SU(2)_L\times U(1)_Y$, they do not directly modify the
electroweak gauge running.  The bino--wino relation responsible for
the small neutral-Higgsino splitting is therefore unchanged at
leading order, whereas the gluino mass relation and the detailed
unification condition can be modified.  
The spectrum and collider phenomenology of the hidden hadrons, as
well as the quantitative modification of gauge coupling unification,
depend on the detailed hidden-sector realization.

The hidden strong dynamics therefore allows the pNGB scale to remain
near the TeV scale even when the supersymmetric mass scale is much
higher.  For a multi-ten-TeV singlet soft mass and a moderately lower
confinement scale, $f_\phi$ and $m_\phi$ can both lie near the TeV
scale while preserving the effective pNGB description used in the
phenomenological analysis.  We next consider the hidden gauge sector
responsible for the dominant invisible mediator width.


\subsection{Dark gauge sector and the invisible width}
\label{subsec:dark-gauge-realization}

We take $S$ to be a singlet under an unbroken dark gauge group
$U(1)_D$, while the hidden quarks carry vectorlike $U(1)_D$
charges.  The resulting mixed anomaly generates the coupling of the
pNGB $\phi$ to dark gauge bosons, providing the invisible mediator
width required in the phenomenological analysis.  The dark gauge
interaction also allows hidden hadrons to annihilate or decay
efficiently, avoiding potentially problematic relic abundances.

For vectorlike dark charges and the $U(1)_\phi$ charge normalization
adopted above, the hidden-quark contribution to the dark anomaly
coefficient is
\begin{equation}
    C_D
    =
    2\sum_i N_i q_{D,i}^2 ,
    \label{eq:CD-hidden}
\end{equation}
where $N_i$ denotes the multiplicity associated with the hidden
confining gauge representation. 
The same charged matter also contributes to the running of the dark
gauge coupling.  Between $\Lambda_H$ and $m_{\rm SUSY}$, where the
hidden quarks are dynamical while their scalar partners are
decoupled, the one-loop running can be written directly in terms of
$C_D$ as
\begin{equation}
    \frac{1}{\alpha_D(m_{\rm SUSY})}
    =
    \frac{1}{\alpha_D(\Lambda_H)}
    -
    \frac{C_D}{3\pi}
    \ln\left(
        \frac{m_{\rm SUSY}}{\Lambda_H}
    \right).
    \label{eq:dark-running}
\end{equation}
The absence of a Landau pole below $m_{\rm SUSY}$ requires
\begin{equation}
    C_D\alpha_D(\Lambda_H)
    <
    \frac{3\pi}
         {\ln(m_{\rm SUSY}/\Lambda_H)} .
    \label{eq:dark-perturbativity}
\end{equation}
For $m_{\rm SUSY}/\Lambda_H\simeq3$, this gives
$C_D\alpha_D(\Lambda_H)\lesssim8.6$.   
Perturbativity generally imposes a stronger condition, whose precise
form depends on $C_D$ and the charged matter content.  Nevertheless,
values of $C_D\alpha_D$ of order a few, as relevant for the
phenomenological analysis, can be realized with suitable
hidden-sector matter content.

The Abelian description need not extend far above the supersymmetric
mass scale.  For example, $U(1)_D$ may arise as the unbroken subgroup
of a non-Abelian gauge group $G_D$,
\begin{equation}
    G_D
    \longrightarrow
    U(1)_D ,
    \label{eq:dark-nonabelian-embedding}
\end{equation}
with the breaking scale near $m_{\rm SUSY}$.  The Abelian running is
then restricted to the relatively short interval above
$\Lambda_H$, while the ultraviolet behavior is governed by the
non-Abelian theory and depends on its matter content.  A sizable
$C_D\alpha_D$ at the confinement scale therefore does not require
the Abelian theory to remain perturbative up to the GUT or Planck
scale. We also note that, depending on the symmetry-breaking pattern, this completion may
also contain stable hidden magnetic monopoles.
Nonperturbative effects, including small instantons, can generate
an additional potential for $\phi$, while a relic monopole
population induces a density-dependent potential through the
Witten effect~\cite{Fan:2021ntg,Kawasaki:2015lpf,Nomura:2015xil}.
Surviving monopoles could also contribute to the dark matter.

The dark gauge interaction also provides annihilation and decay
channels for hidden hadrons.  States carrying $U(1)_D$ charge can
annihilate into dark gauge bosons, while neutral pseudoscalars with
a nonvanishing dark anomaly can decay through
$\pi_H\to A_DA_D$.  The detailed spectrum and decay pattern depend
on the hidden flavor and charge assignments and must be checked in
a fully specified ultraviolet realization.  Subject to this model
dependence, the same $U(1)_D$ structure that generates the invisible
mediator width can also prevent unwanted stable hidden-hadron relics.


\subsection{Origin of the Higgsino mass and axion coupling}
\label{subsec:higgsino-mass-origin}

A remaining model-building issue is the origin of the Higgsino mass
and its dependence on the $U(1)_\phi$ phase.  The same mechanism
should generate the axion--Higgsino coupling
$g_{\phi\chi\chi}\sim\mu/f_\phi$ without inducing an additional
Majorana mass that spoils the small neutral-Higgsino splitting.  We
do not specify a unique ultraviolet realization, but discuss several
possibilities that satisfy these requirements to different degrees.

The most direct possibility is the superpotential interaction
\begin{equation}
    W
    =
    \lambda S H_uH_d ,
    \label{eq:renormalizable-SHH}
\end{equation}
where the normalization
$q_\phi(H_u)+q_\phi(H_d)=1$ adopted above fixes
$q_\phi(S)=-1$.  Once $S$ acquires a vacuum expectation value, this
interaction generates $\mu=\lambda\langle S\rangle$.  Since the
phase of $S$ contributes to the pNGB associated with spontaneous
$U(1)_\phi$ breaking, the Higgsino mass term inherits the
corresponding pNGB dependence and generates the required
pNGB--Higgsino coupling.  In an ordinary supersymmetric singlet
sector, however, the singlino mixes with the neutral Higgsinos
through electroweak-symmetry-breaking terms of order $\lambda v$.
Integrating out the singlino then induces an additional Majorana
contribution to the neutral-Higgsino masses,
which can exceed the small splitting required for inelastic
Higgsino phenomenology. 
Although this contribution can be suppressed by increasing the
singlino mass, decoupling the singlet supermultiplet as a whole can
also modify $\langle S\rangle$ and the low-energy pNGB sector.

One way to avoid this singlino-induced correction while retaining
the interaction in Eq.~\eqref{eq:renormalizable-SHH} is to realize
$S$ as an appropriately constrained superfield.  Suitable
constrained-superfield constructions can remove the singlino as an
independent low-energy degree of freedom while retaining the bosonic
degrees of freedom responsible for $U(1)_\phi$
breaking~\cite{DallAgata:2015zxp,Benakli:2017whb}.
The same interaction can then generate the Higgsino mass and its
associated pNGB dependence without inducing the singlino-mediated
Majorana correction.  The detailed implementation is model dependent,
however, particularly through the auxiliary component of $S$ and the
supersymmetry-breaking interactions induced in the Higgs sector.
We therefore regard such a constrained-superfield realization as a
viable implementation of Eq.~\eqref{eq:renormalizable-SHH}, rather
than as a necessary ingredient of the framework.

 
A second possibility is to generate the Higgsino mass through a
field-dependent Giudice--Masiero interaction~\cite{Giudice:1988yz}.
Representative K\"ahler-potential operators take the form
\begin{equation}
    K =
    \frac{S^\dagger}{\Lambda_1}H_uH_d + {\rm h.c.}
    \qquad {\rm or} \qquad
    K =
    \frac{S^{\dagger 2}}{\Lambda_2^2}H_uH_d + {\rm h.c.},
    \label{eq:GM-SHH}
\end{equation}
where the effective suppression scales absorb the corresponding
dimensionless coefficients.  The two operators correspond to
different $U(1)_\phi$ charge assignments and should be regarded as
distinct realizations.  After supersymmetry and $U(1)_\phi$ breaking,
they can generate an effective Higgsino mass carrying the pNGB phase,
and hence the required axion--Higgsino coupling, without introducing
the direct singlino--Higgsino mixing associated with the
superpotential realization.  The relation between the effective
Higgsino mass and the suppression scales $\Lambda_{1,2}$ depends on
the supersymmetry-breaking properties of the $S$ sector, in particular
on its auxiliary component and the associated soft interactions.

The Giudice--Masiero realizations are subject to additional
consistency conditions that depend on the supersymmetry-breaking
properties of the $S$ sector.  In the tadpole-induced stabilization
considered in Sec.~\ref{subsec:hidden-strong-dynamics}, the operator proportional to
$S^\dagger H_uH_d$ can induce a sizable effective interaction along
the MSSM $D$-flat direction and potentially destabilize the Higgs
potential.  This constraint should be analyzed together with the
heavy-Higgs sector, whose mass is comparable to the singlet soft
mass in the present setup.

A separate issue arises from the large hierarchy
$m_{3/2}\gg m_{\rm SUSY}$.  Tree-level supergravity effects involving
$m_{3/2}$ can generate sizable scalar interactions associated with
either operator in Eq.~\eqref{eq:GM-SHH}.  Their magnitude depends
on the supersymmetry-breaking properties of $S$ and on the ultraviolet
origin of the effective operators, and can become important after the
heavy fields are integrated out.  The viability of these realizations
therefore depends on the resulting scalar potential.

We do not pursue a specific ultraviolet origin of the $\mu$ term
here.  For the low-energy phenomenology, the essential requirement
is that the Higgsino mass carry the appropriate $U(1)_\phi$ phase
while preserving the small neutral-Higgsino splitting.


\subsection{Extended Higgs-sector realization}
\label{subsec:extended-higgs}

An alternative realization can be obtained by extending the MSSM
Higgs sector with an additional pair of doublets, $H'_u$ and $H'_d$.
A simple example is provided by the
$U(1)_\phi$-invariant superpotential
\begin{equation}
    W =
    \mu H_uH_d
    + \lambda S H'_uH_d
    + \lambda' S H'_uH'_d ,
    \label{eq:extended-higgs-W}
\end{equation}
with $|\mu|\gg \lambda\langle S\rangle,\lambda'\langle S\rangle$.
The heavier Higgsino pair is then predominantly composed of the MSSM
Higgsinos, while the lighter pair is predominantly composed of the
additional Higgsinos, with
\begin{equation}
    m_\chi \simeq \lambda'\langle S\rangle ,
    \label{eq:extended-higgs-mchi}
\end{equation}
up to corrections suppressed by $\langle S\rangle/\mu$.  The light
Higgsino mass therefore carries the pNGB phase associated with
$U(1)_\phi$ breaking, generating the axion--Higgsino coupling
required for the annihilation mechanism.

This realization also provides a simple origin for the quasi-Dirac
Higgsino spectrum.  We assume that the neutral components of the
additional Higgs doublets acquire small vacuum expectation values,
\begin{equation}
    v'_u,\;v'_d \ll v_{\rm EW},
    \label{eq:small-additional-vevs}
\end{equation}
where
$v_{\rm EW}^2=v_u^2+v_d^2+v_u^{\prime\,2}+v_d^{\prime\,2}$
denotes the electroweak symmetry-breaking scale.  Such small
additional-Higgs vacuum expectation values can arise from the
Higgs-sector soft terms. 
The gaugino-induced Majorana masses of the additional
neutral Higgsinos are then parametrically suppressed by the small
additional-Higgs vacuum expectation values.  The neutral-state
splitting can consequently remain well below the charged--neutral
splitting without relying on the bino--wino cancellation discussed
in Sec.~6.1.  The electroweak radiative contribution to the
charged--neutral splitting, on the other hand, remains at the
sub-GeV scale. 
The hierarchy $\Delta m_0\ll\Delta m_\pm$ required for the
environment-dependent population effect can therefore be realized
in this setup.

The $U(1)_\phi$ charge assignment has an additional important
consequence.  The superpotential in
Eq.~\eqref{eq:extended-higgs-W} requires
\begin{equation}
    q_\phi(H'_d)=q_\phi(H_d),
    \qquad
    q_\phi(H'_u)=q_\phi(H_u)-q_\phi(S),
    \qquad
    q_\phi(H_u)+q_\phi(H_d)=0.
    \label{eq:extended-higgs-charges}
\end{equation}
This charge assignment differs from that of the minimal MSSM
realization in Sec.~\ref{sec:axion-couplings}, since the explicit
$\mu H_uH_d$ term requires
$q_\phi(H_u)+q_\phi(H_d)=0$.  In the present realization, the
normalization of the pNGB coupling to the light additional Higgsinos
is instead set by the $U(1)_\phi$-breaking mass term
$\lambda' S H'_uH'_d$.
After electroweak symmetry breaking, the physical pNGB must be
orthogonal to the neutral electroweak Goldstone mode.
As a result,
its projection onto the MSSM Higgs directions vanishes for
$v'_{u,d}\to0$.  For small additional-Higgs vacuum expectation
values, the resulting couplings to Standard Model fermions are
parametrically suppressed,
\begin{equation}
    c_f
    =
    {\cal O}\left(
        \frac{v_u^{\prime\,2}}{v_{\rm EW}^2}
    \right),
    \label{eq:extended-higgs-fermion-couplings}
\end{equation}
for $f=u,d,\ell$, where $c_f$ denotes the derivative coupling
defined in Eq.~\eqref{eq:axion-fermion-derivative}.  
The dependence on $v'_u$ alone follows from the identical
$U(1)_\phi$ charges of $H_d$ and $H'_d$.  
The $t\bar t$ channel can therefore be strongly suppressed
independently of $\tan\beta$, while the fermionic contribution to
the high-energy solar-neutrino signal is parametrically reduced.

The gauge-boson couplings have the same low-energy structure as
those discussed in Sec.~\ref{subsec:axion-anomalies}, and are
described by the threshold-matched coefficients
$C_i^{\rm eff}=C_i+C_i^{\rm loop}$.
In the absence of additional heavy colored matter, the physical QCD
coefficient $C_3^{\rm eff}$ is suppressed because the pNGB couplings
to SM quarks are small in the present realization.  The gluonic
decay mode is therefore strongly suppressed unless heavy colored
matter generates a sizable local QCD anomaly contribution.

The light additional Higgsinos, on the other hand, have the same
electroweak quantum numbers as the MSSM Higgsinos and acquire their
mass from $U(1)_\phi$ breaking.  Their near-threshold loops therefore
give unsuppressed contributions to $C_{1,2}^{\rm eff}$ with the same
characteristic electroweak pattern as in
Sec.~\ref{subsec:axion-anomalies}.  The leading visible gauge-boson
modes are consequently
\begin{equation}
    \phi\to W^+W^-,\quad ZZ,\quad Z\gamma,\quad\gamma\gamma .
    \label{eq:extended-higgs-visible-modes}
\end{equation}
These channels remain unsuppressed even when the direct fermionic
modes and the gluonic channel are strongly suppressed.  The
suppression of the fermionic modes reduces their contribution to
the solar-neutrino signal and weakens its dependence on $\tan\beta$,
while the $Z\gamma$ and $\gamma\gamma$ channels retain the Galactic
gamma-ray line signatures.

Apart from the modified mediator couplings, the leading dark-matter
phenomenology developed above applies to the additional Higgsino.
The light state has the same electroweak gauge quantum numbers as
an MSSM Higgsino, and hence the same leading electroweak
annihilation and coannihilation rates. 
In particular, the $WW$ and $ZZ$ channels exhibit the same interference structure
between the standard and axion-mediated amplitudes as in the MSSM
case.  The nearly degenerate neutral and charged states participate
in cosmological reannihilation, while at low temperatures only the
lightest neutral state remains populated.  The population-induced
environmental enhancement is therefore unchanged.  Likewise, the
off-diagonal $Z$ interaction gives the same inelastic solar-capture
and terrestrial-scattering mechanisms.  The suppressed fermionic
mediator decays reduce the high-energy solar-neutrino yield, while
the electroweak gauge-boson channels provide complementary Galactic
gamma-ray signatures.

The additional Higgs doublets modify the electroweak gauge beta
functions and hence the gaugino-mass evolution and the
precision-unification condition.  The gaugino-induced Majorana
masses of the additional neutral Higgsinos are already suppressed
by the small $v'_{u,d}$, so the bino--wino cancellation and the
associated condition $\alpha\simeq-3$ discussed in Sec.~\ref{subsec:mirage-realization}  
are not required in this realization.  A quantitative analysis of
gauge coupling unification and the extended Higgs spectrum depends
on the additional thresholds and is beyond the scope of the present
discussion.

This construction therefore provides an alternative supersymmetric
realization of the low-energy framework developed above.  The
quasi-Dirac structure follows from the small additional-Higgs vacuum
expectation values, which also strongly suppress the pNGB couplings
to SM fermions.  At the same time, the near-threshold additional
Higgsinos generate unsuppressed pNGB couplings to
electroweak gauge bosons through $C_{1,2}^{\rm eff}$.  The resulting
separation between the fermionic and electroweak gauge-boson
couplings allows the solar-neutrino signal to be suppressed while
retaining the Galactic gamma-ray line signatures.


\subsection{Dark radiation}
\label{subsec:dark-radiation}

The massless dark photon responsible for the dominant invisible
mediator width contributes to dark radiation, providing an
independent cosmological consequence of the model.  The dark-radiation
abundance receives a thermal contribution produced while the visible
and dark sectors are in equilibrium, as well as a nonthermal
contribution generated by the late modulus--gravitino cosmology~\cite{Ichikawa:2007jv,Nakayama:2010vs}.

If the dark photon is the only relativistic hidden degree of freedom
remaining after the two sectors decouple, entropy conservation gives
\begin{equation}
    \frac{T_{A'}}{T_\nu}
    =
    \left[
        \frac{g_{*s}(T_{\nu,{\rm dec}})}
             {g_{*s}(T_{\rm dec})}
    \right]^{1/3},
    \label{eq:dark-photon-temperature}
\end{equation}
where $T_{\rm dec}$ denotes the temperature at which the visible and
dark sectors decouple.  The corresponding thermal contribution to
the effective number of neutrino species is~\cite{Nakayama:2010vs}
\begin{equation}
    \Delta N_{\rm eff}^{\rm th}
    =
    \frac{8}{7}
    \left[
        \frac{10.75}{g_{*s}(T_{\rm dec})}
    \right]^{4/3}.
    \label{eq:thermal-delta-neff}
\end{equation}
For the couplings and mass scales relevant to the present scenario,
a parametric estimate of the pNGB-mediated energy-transfer rate
places the decoupling temperature around the electroweak scale.
This corresponds to $g_{*s}(T_{\rm dec})\sim10^2$ and hence to a
thermal contribution $\Delta N_{\rm eff}^{\rm th}$ at the
few-times-$10^{-2}$ level.  For example,
$g_{*s}(T_{\rm dec})\simeq86$ gives
$\Delta N_{\rm eff}^{\rm th}\simeq0.07$.  A precise determination
requires a thermal treatment of the pNGB-mediated energy transfer
across the relevant electroweak thresholds.

The modulus-induced gravitinos provide a second source of dark
radiation.  We define their energy density relative to the
pre-existing radiation immediately before decay by
\begin{equation}
    r_{3/2}
    \equiv
    \left.
    \frac{\rho_{3/2}}{\rho_R}
    \right|_{T=T_{3/2}}
    \simeq
    \frac{4}{3}
    \frac{m_{3/2}Y_{3/2}}{T_{3/2}},
    \label{eq:r-gravitino}
\end{equation}
where the last relation neglects the small difference between the
effective energy and entropy degrees of freedom.  We denote by
$B_H$ the fraction of the gravitino energy initially deposited into
the hidden sector.  Since the primary hidden decay products are
produced far above the confinement scale, their energy is
subsequently redistributed through hidden showering, confinement,
annihilation, and cascade decays.  We parameterize by $\epsilon_D$
the fraction of this energy ultimately transferred to relativistic
dark photons.  Values of $\epsilon_D$ of order unity are possible,
although its precise value depends on the hidden-hadron and
superpartner spectrum.

The resulting nonthermal contribution is given approximately by
\begin{equation}
    \Delta N_{\rm eff}^{3/2}
    \simeq
    \frac{43}{7}
    \left[
        \frac{10.75}{g_{*s}(T_{3/2})}
    \right]^{1/3}
    \frac{
        \epsilon_D B_H r_{3/2}
    }{
        1+(1-B_H)r_{3/2}
    } .
    \label{eq:nonthermal-delta-neff}
\end{equation}
The denominator accounts for the increase in the visible radiation
energy density due to gravitino decay.  At this level of
approximation, the associated entropy release also dilutes the
pre-existing thermal dark-photon population.  Combining the thermal
and nonthermal contributions, the total dark-radiation abundance is
therefore approximately
\begin{equation}
    \Delta N_{\rm eff}
    \simeq
    \frac{\Delta N_{\rm eff}^{\rm th}}
         {1+(1-B_H)r_{3/2}}
    +
    \Delta N_{\rm eff}^{3/2}.
    \label{eq:total-delta-neff}
\end{equation}
In the modulus cosmology considered here, $r_{3/2}$ can be at the
few-percent level while the injected Higgsinos remain in the
reannihilation regime.  If $\epsilon_D B_H$ is not strongly
suppressed, the nonthermal contribution can be comparable to the
thermal component, allowing $\Delta N_{\rm eff}$ to reach the
${\cal O}(0.1)$ level even when gravitinos remain subdominant at
decay.  The precise value depends on the gravitino branching
fractions and the subsequent hidden-sector cascade.

If the visible and dark sectors were once in thermal equilibrium,
the thermal contribution is determined primarily by the decoupling
temperature and depends only weakly on the details of the
hidden-sector cascade.  Dark radiation therefore provides a
complementary cosmological probe of the hidden gauge sector
responsible for the invisible mediator width, independently of the
tentative interpretation of the terrestrial high-recoil event.

\section{Conclusions}
\label{sec:conclusions}

We have studied environment-dependent annihilation of quasi-Dirac
Higgsino dark matter coupled to an axion-like mediator slightly below
the Higgsino-pair threshold. Both subthreshold kinematics and the change
from a coannihilating multiplet to the lightest neutral state enhance
the relative importance of axion-mediated annihilation in the Sun and
Galactic halo compared with cosmological reannihilation following late
nonthermal production. In the degenerate-multiplet limit, the population
effect alone enhances the axion-to-electroweak annihilation ratio by
approximately $12.4$.

In the Sun, capture--annihilation equilibrium fixes the total annihilation
rate, so dominant annihilation into massless dark photons suppresses the
high-energy neutrino signal without suppressing capture. This mechanism
can substantially relax solar-neutrino constraints on Higgsinos
constituting a few percent of the total dark matter abundance, while
preserving an appreciable terrestrial inelastic-scattering rate.
In the Galactic halo, the same low-velocity enhancement instead increases
the annihilation rate. The near-threshold Higgsino loop generates
electroweak continuum and $\gamma\gamma$ and $Z\gamma$ line contributions,
making gamma-ray line searches a particularly promising probe.

We have also discussed supersymmetric realizations based on generalized
mirage mediation and hidden strong dynamics, together with an extended
Higgs-sector alternative. The massless dark photons provide an additional
cosmological signature through their contribution to $\Delta N_{\rm eff}$.

The LZ high-recoil candidate provides a concrete motivation to explore
this framework further. A dedicated fit to the recoil spectrum, including
the high-energy sideband and detector response, is needed to determine
the preferred Higgsino mass, neutral-state splitting, and abundance
fraction. Such an analysis should account for uncertainties in the
high-velocity tail of the Galactic distribution and assess consistency
with solar-neutrino and gamma-ray constraints. Establishing a common
parameter region is an important next step toward a quantitative
Higgsino interpretation of the LZ event. Further high-recoil measurements
and gamma-ray line searches would then provide complementary tests
of this interpretation.

\acknowledgments
This work was supported by the National Research Foundation of Korea
(NRF) grants funded by the Korea government (MSIT)
(RS-2026-25476624, RS-2026-25571203) (K.S.J.).
This work was also supported by JSPS KAKENHI Grant Numbers
25H02165 and 26K00695 (F.T.), by the World Premier International
Research Center Initiative (WPI), MEXT, Japan, and by COST Action
COSMIC WISPers CA21106.

\bibliographystyle{JHEP}
\bibliography{references}

\end{document}